\documentclass[10pt,amsmath,superscriptaddress,amssymb,aps,prd,floatfix,twocolumn]{revtex4-2}

\usepackage[utf8]{inputenc}
\usepackage{epsfig}
\usepackage{graphicx}
\usepackage{mathrsfs}
\usepackage[bookmarks=false,hidelinks]{hyperref}
\usepackage{dcolumn}
\usepackage{bm}
\usepackage{soul}
\usepackage{amsmath}
\usepackage{bbold}
\usepackage{braket}
\usepackage{placeins}

\usepackage{comment}
\usepackage{todonotes}

\begin{document}

\title{Transport coefficients and quasinormal modes in Einstein-dilaton holographic QCD}
\author{Nairy A. Villarreal}
\email{n.villarreal@ufabc.edu.br}
\affiliation {Centro de Ci\^encias Naturais e Humanas, Universidade
Federal do ABC, Avenida dos Estados 5001, 09210-580 Santo Andr\'e, São Paulo,
Brazil}
\affiliation {Laborat\'orio de Astrof\'isica Te\'orica e Observacional, Departamento de Ci\^encias Exatas e Tecnol\'ogicas, Universidade Estadual de Santa Cruz, 45650-000, Ilh\'eus, Bahia, Brazil}

\author{Luis A. H. Mamani}%
\email{luis.mamani@ufrb.edu.br}
\affiliation {Centro de Ci\^encias Exatas e Tecnol\'ogicas, Universidade Federal do Rec\^oncavo da Bahia,\\
Rua Rui Barbosa, 710, 44380-000, Cruz das Almas, Bahia, Brazil}
\affiliation {Laborat\'orio de Astrof\'isica Te\'orica e Observacional, Departamento de Ci\^encias Exatas e Tecnol\'ogicas, Universidade Estadual de Santa Cruz, 45650-000, Ilh\'eus, Bahia, Brazil}

\author{Alfonso Ballon-Bayona}%
\email{aballonb@if.ufrj.br}
\affiliation {Instituto de F\'{i}sica, Universidade
Federal do Rio de Janeiro,\\
Caixa Postal 68528, RJ 21941-972, Brazil.}

\author{Alex S. Miranda}%
\email{asmiranda@uesc.br}
\affiliation {Laborat\'orio de Astrof\'isica Te\'orica e Observacional, Departamento de Ci\^encias Exatas e Tecnol\'ogicas, Universidade Estadual de Santa Cruz, 45650-000, Ilh\'eus, Bahia, Brazil}
 
\author{Vilson T. Zanchin}%
\email{zanchin@ufabc.edu.br}
\affiliation {Centro de Ci\^encias Naturais e Humanas, Universidade
Federal do ABC, Avenida dos Estados 5001, 09210-580 Santo Andr\'e, São Paulo,
Brazil}

\begin{abstract}

In this paper, we investigate the transport coefficients of a strongly coupled plasma in the context of holographic QCD models based on Einstein-dilaton gravity that are compatible with linear confinement at zero temperature. At finite temperature, the holographic model is characterized by an asymptotically anti-de Sitter (AdS) black hole coupled to a scalar field, the dilaton, which is quadratic in the radial direction. The inclusion of the scalar field results in an explicit breaking of the conformal symmetry in the dual field theory. In such systems, the Hawking temperature of the black hole corresponds to the plasma temperature in the dual field theory. We confirm the existence of a minimum temperature $T_{\min}$, above which two distinct classes of black hole solutions emerge: one corresponding to large black holes and the other to small black holes. We calculate some thermodynamic quantities---such as entropy, specific heat, and speed of sound--- and find results that are consistent with similar holographic models. We calculate the quasinormal modes (QNM) of the tensor and vector sectors using the pseudospectral method. In the hydrodynamic regime, we derive the dispersion relation for the vector sector, from which we extract the shear viscosity and the ratio $\eta/s=1/4 \pi$. The bulk viscosity is calculated using the Kubo formula in the scalar sector. Finally, our results for the speed of sound are compared with the Lattice QCD predictions, and our results for the bulk viscosity are compared with those reported by the JETSCAPE collaboration.

\end{abstract}

\maketitle

\section{Introduction}
\label{introd}

It is well established that quarks and gluons are confined at low temperatures and densities. However, at high temperatures and/or densities, these particles briefly remain in a deconfined state known as the Quark-Gluon Plasma (QGP), first observed experimentally at the {\it{Relativistic Heavy Ion Collider}} (RHIC) \cite{adcox2005RHIC, PHOBOS:2004zne, BRAHMS:2004adc, adams2005RHIC}. In the QGP phase, the particles behave collectively, exhibiting characteristics similar to those of a nearly perfect fluid \cite{Natsuume:2014sfa}. Due to this fluid-like behavior, relativistic hydrodynamics emerges as an effective theoretical framework for describing the dynamics of the QGP. In this framework, transport coefficients and dispersion relations naturally arise as fundamental quantities characterizing the dynamic response of the plasma.

The gauge/gravity duality (or AdS/CFT correspondence) \cite{Maldacena:1997re} has proven to be a powerful theoretical framework for computing transport coefficients in strongly coupled gauge theories, such as shear and bulk viscosities. This approach establishes a connection between gravitational perturbations in an asymptotically anti-de Sitter (AdS) spacetime and properties of the dual quantum field theory. The retarded Green's functions of the 4-dimensional finite-temperature field theory can be computed by following the Kovtun–Son–Starinets prescription \cite{Son:2002sd,Kovtun:2005ev,Herzog:2002pc,Policastro:2001yc,Policastro:2002tn}, which involves studying first-order perturbations of a 5-dimensional asymptotically AdS black hole. To ensure causality, one must impose the infalling-wave condition at the event horizon, complemented by a Dirichlet condition at the boundary. One of the most remarkable results derived from this framework is the ratio of shear viscosity ($\eta$) to entropy density ($s$), given by $\eta/ s=1/4\pi$ \cite{Kovtun:2004de,Policastro:2001yc}, valid for a broad class of strongly coupled gauge theories.

In conformal field theories (CFTs), the bulk viscosity ($\zeta$) vanishes, while the speed of sound remains constant. In particular, the speed of sound of a conformal plasma in four dimensions is $1/\sqrt{3}$. When conformal symmetry is broken---as in models that attempt to describe QCD-like theories---these quantities acquire a non-trivial dependence on the temperature. The gauge/gravity duality has become a valuable method for calculating these transport coefficients in 4-dimensional non-conformal plasma and the results can be remarkably similar to those expected for the quark-gluon plasma, see e.g. \cite{Gubser:2008ny,Gubser:2008yx,Ballon-Bayona:2021tzw}.

So far, a considerable number of works have attempted to describe certain properties of QCD at zero and finite temperature using a bottom-up approach usually known as AdS/QCD or holographic QCD; see, for instance, Refs.~\cite{Gubser:2008ny, Gursoy:2010fj, Gursoy:2007cb, Gursoy:2007er, Gursoy:2010fj, Gursoy:2008za, Noronha-Hostler:2015qmd, Yang:2022ixy, Ballon-Bayona:2017sxa} and references therein. Most of these studies employ an Einstein-dilaton theory to describe the gravitational side of the duality. In one scenario the potential of the dilaton field is built using information derived from the phenomenology of the dual field theory while the metric and dilaton field are solved numerically. In another scenario, one constructs the dilaton field or the metric directly, also using results from the dual field theory, and the dilaton potential is reconstructed numerically. In both scenarios the dilaton field deforms the geometry of the AdS spacetime, and this is the 5d gravity dual of the breaking conformal symmetry in the 4d dual field theory. 

In the holographic QCD model of Refs.~\cite{Gubser:2008sz, DeWolfe:2011ts}, the shear and bulk viscosities were calculated for a non-conformal fluid. The computation involves expanding the action to the second order in perturbations, followed by the application of the Kubo formula. This approach has been widely used to obtain transport coefficients in holographic models at finite temperature and density; see, for instance, Refs.~\cite{Rougemont:2015wca, Grefa:2022sav, Gursoy:2009kk} and references therein. The bulk viscosity, calculated via the Kubo formula, introduces a non-arbitrary normalization factor determined through numerical integration. The result obtained using this method contrasts with that derived from the fluid/gravity correspondence in~\cite{Eling:2011ms}, which does not involve any normalization factor. Moreover, the authors of Ref.~\cite{Buchel:2011wx} have shown that both results agree in certain theories at high temperatures. It is also worth mentioning the conjecture for the ratio of the bulk viscosity to the entropy density, $\zeta/\eta \geq 2(1/3 - v_{s}^{2})$, proposed in Ref.~\cite{Buchel:2007mf}. However, evidence of violations of this inequality was obtained in Ref.~\cite{Buchel:2011uj}.

In potential reconstruction holographic models, the dilaton field profile is identified as essential for capturing certain properties of the dual field theory. Karch, Katz, Son and Stephanov \cite{Karch:2006pv} demonstrated that a dilaton that is quadratic in the radial coordinate far from the boundary leads to asymptotically linear Regge trajectories for hadrons. Inspired by these insights, several holographic models~\cite{Li:2014dsa, He:2013qq,Ballon-Bayona:2017sxa, Mamani:2020pks, Miranda:2009uw, Ballon-Bayona:2020xls, Ballon-Bayona:2021tzw} have been developed to investigate various phenomena based on the available experimental data and on the results of Lattice QCD. The exact quadratic dilaton profile, originally proposed in the soft-wall model~\cite{Karch:2006pv}, was later extended to Einstein-dilaton theories~\cite{Gursoy:2007er, Andreev:2007vn,Li:2013oda} and studied at finite temperature and density in Ref.~\cite{Ballon-Bayona:2020xls}. The warp factor in this model is determined analytically from the Einstein equations, whereas the horizon function is given in terms of integrals of the warp factor, as described in Ref.~\cite{Ballon-Bayona:2020xls}. The thermodynamic properties of the semi-analytical model with an exact quadratic dilaton profile are consistent with the results of more sophisticated holographic approaches \cite{Gursoy:2007cb, Gursoy:2007er, Gursoy:2010fj, Ballon-Bayona:2021tzw}.

Motivated by the findings presented in Ref.~\cite{Ballon-Bayona:2020xls}, here we investigate the shear and bulk viscosities within this holographic model. To calculate the shear viscosity, we adopt the method outlined in Refs.~\cite{Kovtun:2005ev, Mas:2007ng, Springer:2008js}, which involves solving the perturbation equations in the hydrodynamic limit to obtain the dispersion relation. The bulk viscosity is calculated using the Kubo formula, following the approach presented in Ref.~\cite{DeWolfe:2011ts}. We also compare our results with those obtained by the JETSCAPE Collaboration \cite{JETSCAPE:2020shq}. Additionally, we compute the quasinormal frequencies in the tensor and vector sectors by employing the pseudospectral method, initially implemented in the \textit{Mathematica} package from Ref.~\cite{Jansen:2017oag}. For cross-validation, we further evaluate the quasinormal frequencies using the numerical codes developed in Ref.~\cite{Mamani:2022qnf}.

This article is organized as follows. In Section~\ref{Sec:holographicmodel}, we review the black hole thermodynamics and present results for relevant physical quantities, including temperature, entropy, and the speed of sound. The latter is compared against the Lattice QCD results. In Section~\ref{Sec:Perturbation}, we introduce the linear perturbation theory and define the gauge-invariant quantities employed in this model. Sections~\ref{Subsec:TThydro} and~\ref{Subsec:Vecthydro} detail the general procedures for obtaining solutions in the tensor (transverse and traceless) and vector (shear) sectors, respectively. In Section~\ref{subsec:bulkviscosity}, we compute the bulk viscosity using the Kubo formula. In Section~\ref{Sec:Comparison}, we compare our results for the bulk viscosity and speed of sound with the data reported by the JETSCAPE Collaboration and the results from Lattice QCD, respectively. Finally, Section~\ref{Sec:conclusion} presents our conclusions. Additional technical details are provided in Appendices~\ref{app:AsymptoticAnalysis}, \ref{app:Invariant}, and~\ref{app:KuboFormula}.

\section{The holographic model} \label{Sec:holographicmodel}

\subsection{Setting the stage}

The holographic model considered here is derived from the following five-dimensional action:
\begin{equation}\label{eq:action}
    S=M_{p}^{3}N_c^2 \int d^{5}x\sqrt{-g}\left(R-\frac{4}{3}\partial^{\mu}\phi\partial_{\mu}\phi+V(\phi)\right), 
\end{equation}
where $g$ is the determinant of the metric, $R$ is the Ricci scalar, $\phi$ is the scalar field (dilaton),
$V(\phi)$ is the dilaton potential, $M_{p}$ is the 5d Planck mass, $N_c$ is a parameter that will be interpreted in four dimensions as the number of colors. For more details of this action see \cite{Gursoy:2007cb,Gursoy:2007er}. The Greek indices ($\mu, \,\nu,\, \rho,\, \dots$) run from $0$ to $4$.

The Einstein equations derived from the action can be written in the Ricci form,
\begin{equation}\label{Eq:EinsteinEquation0}
    R_{\mu \nu}-\frac{4}{3}\partial_{\mu} \phi\partial_{\nu} \phi+\frac{1}{3}g_{\mu \nu}V(\phi)=0,
\end{equation}
where $g_{\mu \nu}$ is the metric, the Ricci tensor is given by
$R_{\mu \nu}=R^{\sigma}_{\mu \sigma \nu}=\partial_{\sigma}\Gamma^{\sigma}_{\nu \mu}-\partial_{\nu}\Gamma^{\sigma}_{\sigma\mu}+\Gamma^{\sigma}_{\sigma \rho}\Gamma^{\rho}_{\nu \mu}-\Gamma^{\sigma}_{\nu \rho}\Gamma^{\rho}_{\sigma \mu}$, and the Christoffel symbol is defined by $\Gamma^{\sigma}_{\mu \nu}=\frac{1}{2}g^{\sigma \rho}(\partial_{\mu}g_{\rho\nu}+\partial_{\nu}g_{\rho\mu}-\partial_{\rho}g_{\mu \nu})$.
Additionally, the Klein-Gordon equation takes the form:
\begin{equation}\label{Eq:KleinGordon0}
    \frac{8}{3}\frac{1}{\sqrt{-g}}\partial_{\mu}\left[\sqrt{-g}g^{\mu \nu}\partial_{\nu}\phi\right]+\frac{dV(\phi)}{d\phi}=0.
\end{equation}

From here on, we consider the following 5-dimensional metric ansatz:
\begin{equation}\label{Eq:Metric}
ds^2=-\frac{f(z)}{\zeta_1^2(z)}\,dt^2+\frac{1}{f(z)\zeta_2^2(z)}\,dz^2+\,\frac{1}{\zeta_1^2(z)}\,d\vec{x}^{2},
\end{equation}
where $t$ is the time, $z$ is the holographic radial coordinate, defined in the range $0\leq z < \infty$,
and the quantity $d\vec{x}^{\,2}= dx_i dx^i$ denotes the Euclidean metric in a 3-dimensional space,
with Latin indices $(i,\, j, \, k,\, \cdots$) running from $1$ to $3$.

In Eq.~\eqref{Eq:Metric}, $f(z)$ is the horizon function
(also known as the blackening function) that vanishes at the horizon $z_h$, i.e., $f(z_{h})=0$.
In these coordinates, the AdS boundary is located at $z=0$, where the horizon function assumes the value $f(0)=1$.

Substituting the metric \eqref{Eq:Metric} into Eqs.~\eqref{Eq:EinsteinEquation0} and \eqref{Eq:KleinGordon0}, one obtains: 
\begin{align}\label{Eq:MotionZero}
&  \frac{\partial_z^2\zeta_{1}}{\zeta_{1}} + \frac{\partial_z\zeta_{1}}{\zeta_{1}}\frac{\partial_z\zeta_{2}}{\zeta_{2}}
- \left(\frac{\partial_z\zeta_{1}}{\zeta_{1}}\right)^2 - \frac{4}{9}\left(\partial_z\phi\right)^2 = 0,\\
&  V(\phi) + \frac{3 \zeta_2^2 \partial_z f\partial_z\zeta_1}{\zeta_1} - \frac{12 f\zeta_2^2 \left(\partial_z\zeta_1\right)^2}{\zeta_1^2}
+ \frac{4}{3} f\zeta_2^2 \left(\partial_z\phi\right)^2 = 0,\label{Eq:DilatonPotencial}\\
&  \frac{\partial_z^2 f}{\partial_z f} - \frac{4 \partial_z\zeta_{1}}{\zeta_1} + \frac{\partial_z\zeta_2}{\zeta_2} = 0,\\
&   \partial_{\phi}V(\phi) + \frac{8f\zeta_{2}^{2}}{3\partial_z f}\left(\partial_z f\partial_z^2\phi
- f\partial_z^2\ln f\partial_z\phi\right) = 0.\label{Eq:Phiorderzero}
\end{align}

Although most calculations are performed assuming $\zeta_{1}(z)=\zeta_{2}(z)$, we present the general form of the equations here,
as they are useful for studying perturbations and facilitate comparison with the different notation found in the literature.

\subsection{The dilaton ansatz and other simplifying assumptions}

To solve the system of equations \eqref{Eq:MotionZero}--\eqref{Eq:Phiorderzero},
in addition to imposing the condition
\begin{equation}
      \zeta_{2}(z)=\zeta_{1}(z)\,,
\label{Eq:zeta2}
\end{equation}
we adopt a strategy similar to the one employed in Refs.~\cite{Li:2013oda, Ballon-Bayona:2020xls},
i.e., we fix the quadratic dilaton profile 
\begin{equation}
\phi(z)=c\,z^2, \label{eq:dilaton}
\end{equation}
where $c$ is a constant parameter that will be fixed later.
This choice leads to an exact solution for the warp factor $\zeta_{1}(z)$, with the horizon function $f(z)$ being determined numerically.
The potential $V(\phi)$ is obtained by substituting the solutions for the warp factor and the horizon function into Eq.~\eqref{Eq:DilatonPotencial}.
This procedure for solving the coupled equations is known as the potential reconstruction method and has been extensively applied in the literature; see, for example, Refs.~\cite{Li:2013oda, Li:2014hja, Mamani:2020pks, He:2013qq, Arefeva:2020vae, Mamani:2022qnf} and references therein.
The quadratic dilaton \eqref{eq:dilaton}, combined with assumption \eqref{Eq:zeta2}, enables us to find semi-analytical solutions for the thermodynamic quantities of the asymptotically AdS black hole, including the Hawking temperature, entropy density, free energy, specific heat, and the speed of sound.

The assumption $\zeta_2=\zeta_1$ allows us to rewrite the system of equations \eqref{Eq:MotionZero}--\eqref{Eq:Phiorderzero} in the form:
\begin{align}
&  \frac{\partial_z^2\zeta_{1}}{\zeta_{1}} - \frac{4}{9}\left(\partial_z\phi\right)^2=0\,,
    \label{Eq:Zeta}\\
&  \frac{\partial_z^2 f}{\partial_z f} - 3\frac{\partial_z\zeta_{1}}{\zeta_1} = 0\,,
    \label{Eq:horizonfuntion}\\
&  3\partial_z^2\zeta_1 - 12\frac{\left(\partial_z\zeta_{1}\right)^{2}}{\zeta_{1}} + 3\frac{\partial_z f\partial_z\zeta_{1}}{f}
    + \frac{V(\phi)}{\zeta_{1}f} = 0\,,
    \label{Eq:VPotencial}\\
&   \partial_{\phi}V(\phi) + \frac{8f\zeta_{1}^{2}}{3\partial_z f}\left(\partial_z f\partial_z^2\phi
- f\partial_z^2\ln f\partial_z\phi\right) = 0\, .
    \label{Eq:DerivativePotencialZero}
\end{align}
Equation \eqref{Eq:Zeta} admits the analytic solution:
\begin{equation}\label{Eq:ZetaAna}
\zeta_{1}(z)=z~ {}_0F_{1}\left[\frac{5}{4},\frac{c^2z^4}{9}\right],
\end{equation}
where $_0F_1[a,x]$ denotes the hypergeometric function. This function satisfies the boundary condition
$\zeta_{1}(z)=z$ as $z\rightarrow 0$, ensuring that the metric is asymptotically AdS, consistent with the solution obtained
in the absence of the dilaton field.
The condition $c=0$, which corresponds to turning off the dilaton field, recovers the conformal plasma description.
The asymptotic behavior of the warp factor $\zeta_{1}(z)$ near the boundary is given by
\begin{equation}\label{Eq:ZetaAsyntotic}
  \zeta_{1}(z)\simeq\, z\left(1+\frac{4}{45}c^2z^4 +\frac{8 c^4 z^8}{3645} + \dots \right). 
\end{equation}
Consequently, $\zeta_{1}(z)$ approaches zero as $z \rightarrow 0$.

\subsection{Thermodynamics}

The thermodynamic properties of this holographic model with a quadratic dilaton field profile and finite
density were investigated in Ref.~\cite{Ballon-Bayona:2020xls}. The results of this section align with
those reported in that work in the zero-density limit. Here, we present the thermodynamic quantities
relevant to the analysis in the next section, such as the temperature, entropy density, and speed of sound.

\subsubsection{Temperature}

First, we calculate the black hole temperature, given by the relation
\begin{equation}
T=-\frac{1}{4\pi}\frac{df(z)}{dz}\bigg|_{z_{h}}. \label{Eq:temp}
\end{equation}
This is the Hawking temperature, which depends on $z_h$, and is interpreted as the temperature of the dual field theory, in this case, a non-conformal theory.

In the present model, the derivative of the horizon function comes from the solution of Eq.~\eqref{Eq:horizonfuntion}.
To solve Eq.~\eqref{Eq:horizonfuntion}, appropriate boundary conditions for $f(z)$ must be specified. As usual, these conditions are $f(z_h)=0$ at the event horizon ($z=z_h$) and $f(0)=1$ at the boundary ($z=0$).
Then, we get
\begin{equation}\label{Eq:horizontsolution}
f(z)=1-C_h\int_{0}^{z} \zeta_1^3(\tilde{z})d\tilde{z}\,,
\end{equation}
where $C_h$ is given by 
\begin{equation}\label{Eq:Ch}
C_h ^{-1}={\int_{0}^{z_h} \zeta_1^3(\tilde{z})d\tilde{z}}\, .
\end{equation}

It is worth mentioning that the blackening function, temperature, and other thermodynamic quantities are obtained by means of numerical methods.
Nevertheless, the asymptotic behavior of these quantities can be derived analytically at small $z_h$ (corresponding to very large black holes), see Appendix \ref{app:AsymptoticAnalysis} for further details.

\begin{figure}[h!]
\centering
\includegraphics[width=7cm]{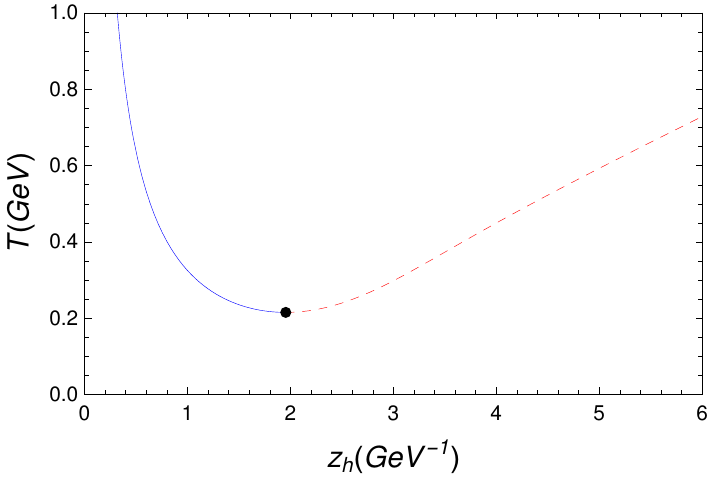}
\caption{Temperature as a function of the horizon position $z_{h}$. The black dot marks the position $z_h=z_{min}$ at which the temperature
attains its minimum value $T_{min}$, considering $c=0.4\,\text{GeV}^{2}$. The solid blue line represents large black holes, while the dashed red
line corresponds to small black holes.}
\label{Fig:Temperature}
\end{figure}

Figure \ref{Fig:Temperature} shows the behavior of the temperature as a function of the horizon position $z_{h}$
for a dilaton field with $c=0.4\,\text{GeV}^{2}$ [see Eq.~\eqref{eq:dilaton}]. 
The non-monotonic behavior of the temperature leads to two distinct branches of AdS black hole configurations. For $z_h<z_{min}$,
corresponding to large black holes (solid blue line), the temperature decreases as $z_h$ increases, indicating locally stable
thermodynamic solutions. Conversely, for $z_h>z_{min}$, in the small black hole regime (red dashed line), the temperature
increases with $z_h$, signaling thermodynamically unstable configurations.

\subsubsection{Entropy and specific heat}

The entropy of the black hole is computed via the Bekenstein–Hawking formula, i.e.,
\begin{equation}
    S=\frac{A}{4G_5}=\frac{4\pi M_p^3 N_c^2 V_3}{\zeta_{1h}^3},
\end{equation}
where $\zeta_{1h}$ represents the value of the function $\zeta_1(z)$ at $z=z_h$,
$A$ is the area of the event horizon (which is a surface of codimension one) and $V_3$ is the three-dimensional volume of a spacelike section $t=\text{constant}$,
$z=\text{constant}$.
Thus, the entropy density, $s=S/V_3$,
can be written as
\begin{equation}
    s=\frac{4 \pi M_p^3 N_c^2}{\zeta_{1h}^3}\,.
    \label{Eq:Entropy}
\end{equation}
The specific heat is defined as 
\begin{equation}
C_{V}=T\frac{ds}{dT}= s \frac{d \log s}{d \log T}\,. 
\label{Eq:CV}
\end{equation}
This quantity can be calculated numerically. 

According to the first and second laws of black hole thermodynamics, the event horizon area $A$ and temperature $T$ determine
the thermal properties of the system. The specific heat is particularly important for providing insight into the system's local stability.
Using relations \eqref{Eq:temp} and \eqref{Eq:Entropy}, the stability condition $C_V>0$ translates into $\left(dA/dz_h\right)\left(dz_h/dT\right)> 0$.
Hence, given that $dA/dz_h<0$ (as expected for asymptotically AdS black holes), this stability condition ($C_V>0$) requires $dT/dz_h<0$.
In this situation, the black hole can reach and maintain thermal equilibrium.

We observe from Fig.~\ref{Fig:Temperature} that $dT/dz_h$ is negative in the region $z_{h}<z_{min}$, i.e.,
for large black holes, which results in positive specific heat values, meaning that the corresponding systems are
considered thermodynamically stable. In the region $z_{h}>z_{min}$, i.e., for small black holes, one has $dT/dz_h>0$,
which results in negative specific heat values; thus, the corresponding systems are considered thermodynamically unstable.

\subsubsection{Free energy and speed of sound}

The free energy can be obtained from the prescription proposed in \cite{Gursoy:2008za},
\begin{equation}\label{Eq:FreeEnergy}
    F= - \int_{\infty}^{z_h}s(\tilde{z}_h)\frac{dT}{d\tilde{z}_h}d\tilde{z}_h\,.
\end{equation}
In this definition, the condition of vanishing free energy for very small black holes, $F(z_h\rightarrow \infty)=0$,
is implicitly imposed. This means that the free energy of very small asymptotically AdS black holes, which coincides with the free energy of the thermal gas, is set to zero. 

The trace of the energy-momentum tensor at finite temperature is given by $\braket{T^{\mu}_{\phantom{\mu}\mu}}=\epsilon-3 P$,
where $\epsilon = F + T s$ is the energy density and $P=-F$ is the pressure.

The speed of sound $v_s$ is defined as
\begin{equation}\label{Eq:SpeedSoundTherm}
    v_{s}^2=\frac{s}{C_{V}} , \qquad \text{or} \qquad v_{s}^2=\frac{d \log T}{d \log s}.
\end{equation}
It is useful to rewrite the speed of sound in terms of $z_h$. The expression $v_s^2= v_{s}^2(z_h)$
can be written as
\begin{equation}\label{eq:vs2of zh}
v_{s}^2=\frac{s}{T}\frac{dT}{dz_{h}}\left(\frac{ds}{dz_{h}}\right)^{-1},
\end{equation}
while the temperature can be represented as
\begin{equation}
    T=\frac{1}{4\pi}C_{h}  \zeta_{1h}^{3}, \label{eq:temp-zh}
\end{equation}
where $C_{h}$ is given by Eq.~\eqref{Eq:Ch}. 
Substituting Eq.~\eqref{Eq:Entropy} and Eq.~\eqref{eq:temp-zh} into Eq.~\eqref{eq:vs2of zh} yields the following result,
\begin{equation}\label{Eq:SpeedTherT}
    v_{s}^{2}=-1+\frac{C_{h}\zeta_{1h}^{4}}{3\,d\zeta_{1h}/dz_{h}} .
\end{equation}

We now examine the asymptotic behavior of the speed of sound in the limit of large black holes and for small dilaton fields,
i.e., for $z_h\rightarrow 0$ and $c\rightarrow 0$.  Using the previous relations, we obtain
\begin{equation}\label{Eq:velocidadesomc0}
v_{s}^{2}\approx \frac{1}{3}-\frac{8 c^2 z_h^4}{27}+\frac{464 c^4 z_{h}^8}{3645}+\cdots\, .
\end{equation}
This approximate relation explicitly shows the conformal value $1/3$ in the sound speed and a term proportional
to $c^{2}$, which is associated with the dilaton. 

\begin{figure}[ht!]
    \centering
    \includegraphics[width=7cm]{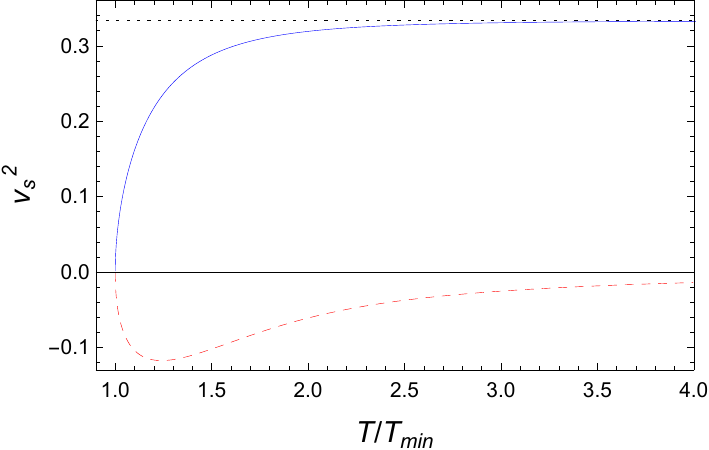}
    \caption{The speed of sound squared as a function of the temperature for $c=0.4\,\text{GeV}^{2}$.
    The blue solid line represents large black holes, while the red dashed line represents small black holes.
    The horizontal black dotted line corresponds to the conformal value $1/3$.}
    \label{Fig:CsT}
\end{figure}

We can also evaluate numerically the speed of sound as a function of the temperature. This is shown in Fig.~\ref{Fig:CsT}. 
This figure shows the two branches of $v_{s}^2(T)$ from Eq.~\eqref{Eq:SpeedSoundTherm} for $c=0.4\,\text{GeV}^{2}$.
A sign change occurs at $T_{min}$, causing $v_{s}^2$ to become negative along one of the branches (red dashed line).
The positive range of $v_{s}^2$ (solid blue line) corresponds to positive specific heat, characteristic of the
large black hole regime. In this regime, for $T\rightarrow \infty$ the asymptotic behavior approaches the conformal value
$v_{s}^2=1/3$. This asymptotic behavior of the speed of sound can be interpreted as suggesting conformal symmetry restoration.
In the small black hole regime, the speed of sound squared ($v_s^2$) is negative, which corresponds to negative values of the specific heat, indicating an unstable phase. We will compare our numerical results for the speed of sound against the Lattice QCD results in section \ref{Sec:Comparison}.
\section{Linear perturbations of the Einstein-Dilaton system}
\label{Sec:Perturbation}

\subsection{The gauge invariant perturbation fields}
\label{Sec:gaugeInvQuantities}

We introduce perturbations into the Einstein-dilaton equations \eqref{Eq:EinsteinEquation0} and \eqref{Eq:KleinGordon0}
by considering metric and dilaton perturbations of the form
\begin{align}
    {g}_{\mu \nu}=&\, \tilde g_{\mu \nu}(z)+h_{\mu \nu}\\
    \Phi=&\,\phi(z)+\chi, \label{Eq:dilatonpert}
\end{align}
where $\tilde g_{\mu \nu}(z)$ stands for the background metric \eqref{Eq:Metric}, $\phi = \phi(z)$ is the background (unperturbed) dilaton field, while the quantities $h_{\mu \nu}$ and $\chi$ represent perturbations to the metric
and scalar field, respectively. The equations of motion for the background quantities $\tilde g_{\mu \nu}(z)$ and $\phi(z)$ presented in
Section \ref{Sec:holographicmodel} are used recursively to simplify the equations for the perturbations, which are
linearized in the perturbation fields. 

Note that we consider perturbations around the background fields that preserve the symmetry transformations of the plane $x^1-x^2$. Therefore, the perturbation fields will not depend on these two coordinates, i.e., $h_{\mu \nu}= h_{\mu \nu}(t,z, x^3)$ and $ \chi=\chi(t,z, x^3)$. We will classify these perturbations according to their transformation properties under this symmetry group into tensor, vector, and scalar sectors.

By Fourier decomposing the metric and dilaton perturbations, we have
\begin{equation}\begin{aligned}
& h_{\mu \nu}(t,z, x^3)=\exp\left[-i \omega t +i q x^{3}\right] h_{\mu \nu}(\omega, q, z),\\
& \chi(t,z, x^3)=\exp\left[-i \omega t +i q x^{3}\right] \chi(\omega, q, z).
    \end{aligned} \label{eq:fourier}
\end{equation}

As is well known, relativistic gauge freedom allows for arbitrary changes of coordinates. For instance, under an infinitesimal coordinate
transformation denoted by $\varepsilon^{\mu}(t,z, x^3)$, the metric and dilaton perturbations transform according to
$\delta g_{\mu \nu}=\varepsilon^{\rho}\partial_{\rho}g_{\mu \nu}+(\partial_{\mu}\varepsilon^{\rho})g_{\rho \nu}+(\partial_{\nu}\varepsilon^ {\rho})g_{\rho \mu}$ and $ 
     \delta\phi=\varepsilon^{\mu}\partial_{\mu}\phi$.
To preserve the symmetry group, the infinitesimal
transformations are also assumed to take the form $\varepsilon^{\mu}(t,z,x^3)=\exp[-i\omega t+iqx^{3}]\varepsilon^{\mu}(z)$.  The gauge symmetry can be used to impose the radial gauge condition, namely $h_{zz}=h_{zt}=h_{zi}=0$ where $i$ denotes the spatial direction, i.e. $i=1,2,3$. 
For the other components, we define the rescaled fields
\begin{equation}
    H_{tt}=\frac{\zeta_{1}^{2}}{f}h_{tt},  \quad H_{t i}=\zeta_{1}^{2} h_{t i},\quad H_{i j}=\zeta_{1}^{2}h_{i j}.
    \label{Eq:Hsdefine}
\end{equation}

Systems related by gauge transformations are equivalent and therefore it is possible to introduce gauge-invariant variables, denoted by $Z_{p}$, where $p$ labels each sector.
Further details on the definition of gauge-invariant variables $Z_{p}=Z_p(z)$ are provided in Appendix \ref{app:Invariant}. 
These variables satisfy second-order linear differential equations, which we call master equations. 

In general, for an asymptotically AdS black hole, the solution to the master equation is given by a combination of two
independent functions representing ingoing (infalling) and outgoing waves.
Imposing the physical condition that waves only fall into the black hole horizon (infalling-wave boundary condition), we select only
the ingoing-wave mode. The other condition is the Dirichlet condition at the AdS boundary. The solution obeying the infalling-wave condition at the horizon can be written as 
\begin{equation}
    Z^{inf}_p(\omega, q, z)=\mathcal{A}\psi_1(z)+\mathcal{B}\psi_2(z),
    \label{zin}
\end{equation}
where $\psi_1(z)$ and $\psi_2(z)$ are the non-normalizable and normalizable solutions near the AdS boundary, respectively. 
The coefficients $\mathcal{A}$ and $\mathcal{B}$ are generally dependent on the frequency $\omega$ and the momentum $q$. 

In the Lorentzian prescription of AdS/CFT \cite{Son:2002sd,Kovtun:2005ev,Herzog:2002pc}, the two-point correlation functions
of an operator $\mathcal{O}$ in the dual field theory are given, up to a multiplicative factor, by the ratio between the
coefficients $\mathcal{B}$ and $\mathcal{A}$:
\begin{equation}
    \braket{\mathcal{O}\mathcal{O}}_{}\sim \frac{\mathcal{B}}{\mathcal{A}}.
\end{equation}
The poles of the retarded correlation functions appear as zeros of the coefficient $\mathcal{A}$.
From the gravitational point of view, the poles of the 4-dimensional correlation functions correspond to the 5-dimensional quasinormal modes (QNMs) of the black hole. For the tensor and vector sectors,
we obtain the dispersion relations associated with QNMs in the hydrodynamic regime ($\omega/T, q/T \ll 1$)
and then, by direct comparison with fluid dynamics, the transport coefficients are identified.
In the scalar sector one can follow the approach presented in Refs.~\cite{Springer:2008js,Mas:2007ng,Abbasi:2020ykq, Kovtun:2005ev},
and obtain two coupled master equations. However, these equations are difficult to solve directly for the QNMs.
In this work, in order to obtain the bulk viscosity, we will use a different method in the scalar sector.
The method is based on the Kubo formula, that relates the transport coefficients of a fluid to the imaginary part of the retarded Green's function in the zero-frequency limit. There are two main ways to obtain the retarded Green's function.
In Ref.~\cite{Gubser:2008sz}, the gravitational action is expanded up to second order in the perturbations to
extract the retarded Green's functions associated with the stress-energy tensor components. For a recent application
of this approach, see Ref.~\cite{Ballon-Bayona:2021tzw}. As an alternative,
in Ref.~\cite{DeWolfe:2011ts} the transport coefficients are extracted from the master equation using Abel's identity. We follow this latter method to evaluate the bulk viscosity, as it avoids the need for an explicit second-order expansion of the action and is better adapted to the characteristics of the perturbation equations under consideration.

\subsection{Setting the stage for calculating the QNMs}\label{Sec:QNMgeneral}

In the holographic model under consideration, the conformal symmetry is broken by the coupling between the dilaton
field and the metric. We follow the approach presented in Refs.~\cite{Kovtun:2005ev, Jansen:2017oag} to compute the
quasinormal modes of the tensor and vector sectors, both analytically and numerically. In this subsection,
we describe the methods and tools used to obtain the QNMs.

The master equations can be expressed in terms of the coordinate $z$, as in Ref.~\cite{Springer:2008js}.
In this work, however, we introduce a dimensionless variable $u$ and normalize the relevant quantities by the event horizon position $z_{h}$:
\begin{equation}
u=\frac{z}{z_h}, \quad \mathfrak{w}=z_{h}\omega, \quad \mathfrak{q}=z_{h}q, \quad \phi_{h}=z_{h}^2 c.
\label{Eq:normalization}
\end{equation}
With these definitions, the master equations become dimensionless, and the event horizon is located at
$u=1$.
The Hawking temperature is given by
\begin{equation}
    T=-\frac{f'_{h}}{4\pi z_{h}},
\end{equation}
where the prime symbol indicates differentiation with respect to $u$
and $f'_{h} = df/du|_{u=1}$.
 
We propose a solution to the master differential equations in the form 
\begin{equation}
Z_{p}(u)=f(u)^{\beta} F_p(u),
\label{GeneralForm}
\end{equation}
where $F_p(u)$ is a regular function at the horizon. To calculate $\beta$, we substitute the ansatz
$Z_{p}(u)=f(u)^{\beta}$ into the master differential equation for $Z_p$. Analyzing this equation near
the event horizon ($u\rightarrow 1$), the leading terms yield an indicial equation for $\beta$:
\begin{equation}
    \beta =\pm \frac{i \mathfrak{w}}{f'_{h}},
    \label{beta}
\end{equation}
corresponding to the ingoing and outgoing waves near the black hole horizon. These leading terms can be written as $\exp(\pm i \mathfrak{w} u^*)$, where we have introduced the tortoise coordinate $u^*$ related to $u$ by $du^*= du/f(u)$.
We must impose the physical condition of purely infalling waves at the horizon. Given that $f'_{h} < 0$,
the choice of positive sign for $\beta$ in \eqref{beta} corresponds to the ingoing wave at the event horizon.
Therefore, the solution for $Z_p$ satisfying the ingoing-wave boundary condition near the horizon takes the form
\begin{equation}\label{Eq:GeneralSolutionZ}
    Z_p(u)=f(u)^{i \mathfrak{w}/f'_{h}} F_p(u).
\end{equation}

The function $F_p(u)$ can be calculated using various methods. One of the proposed methods involves a series expansion of $F_p(u)$
in terms of the frequency and wavenumber. This can be done by replacing $\mathfrak{w}= \lambda \mathfrak{w}$
and $\mathfrak{q}=\lambda \mathfrak{q}$, where $\lambda$ is a parameter that controls the series expansion, and writing
$F_{p}(u)=\sum_{n=0}^{\infty} \lambda^n F_{p}^{n}(u)$. With this, second-order differential equations are obtained
for each order in $\lambda$, which can potentially be solved analytically. The regularity condition at the event
horizon is then used to fix one of the integration constants for each $F_{p}^{n}(u)$. Finally, the Dirichlet condition
is imposed at the boundary to identify the dispersion relations in the hydrodynamic limit, i.e.,
\begin{equation}
    Z_p(u)\Big|_{u=0}=0.
\end{equation}

Another method involves expanding the background functions $f(u)$, $\phi(u)$ and $\zeta_{1}(u)$ near the event horizon ($u=1$).
One can then analyze the behavior of $F_p(u)$ near the horizon, by assuming a leading behavior like $F_{p}(u)\propto (1-u)^n$,
that satisfies regularity at the horizon. Subsequently, the Dirichlet condition at the boundary ($u=0$) is used to identify the dispersion relations.
The results obtained using these two methods in the tensor and vector sectors are identical, but the details are not presented here. Here, we present the first method, where the small frequency and momentum expansion allows systematic and analytic extraction of transport coefficients.

To calculate the QNMs numerically, we employ the pseudo-spectral method, which has been applied to calculate quasinormal frequencies
in various contexts; see, for instance, Refs.~\cite{Jansen:2017oag, Mamani:2022akq}, and references therein. We implement the
pseudo-spectral method in two different ways: the first follows the procedure outlined in Ref.~\cite{Jansen:2017oag},
while the second is based on the analysis presented in Ref.~\cite{Mamani:2022akq}. The second method is used as a consistent check for the numerical results.

Ref.~\cite{Jansen:2017oag} provides the ``QNMspectral'' package, which enables the computation of quasinormal modes using
the pseudo-spectral method implemented in {\textit{Mathematica}}. In this method, each continuous variable in the differential
equations is discretized on a set of points known as the grid, which is constructed using Chebyshev polynomials.
The number of grid points and the computational precision are adjustable parameters. The algorithm has two main functions:
\texttt{GetAccurateModes}, which automatically classifies the true frequencies, and \texttt{GetModes}. The \texttt{GetModes}
function allows computations on a numerically defined background---as is the case in our analysis---but requires a procedure
to identify the true frequencies. For this purpose, we perform computations using two different grid sizes and precision levels
in order to detect spurious numerical results, also referred to as numerical artifacts in Ref.~\cite{Jansen:2017oag}. 
 
This package requires the master equations as input, and it is recommended that they be expressed in
Eddington–Finkelstein (E-F) coordinates so that the equations exhibit an odd dependence on the frequency.
In the present case, it suffices to replace the time coordinate $t$ by the advanced time, while the spacelike
coordinate $u$ (or $z$) remains unchanged, i.e., $dt \to dt - du^*$, where $du^*= du/f(u)$ defines the tortoise
coordinate $u^*=u^*(u)$.  Under the Fourier transform of any perturbation function, this change in time coordinates
is equivalent to replacing the original function $Z_p(u)$ with $ \exp[i\mathfrak{w}u^*] \hat{Z}_p(u)$.

The master equations must satisfy the regularity condition at the event horizon and admit a normalizable
solution at the AdS boundary. The regularity condition at the horizon is obtained by assuming a solution
of the form $Z_{p}\propto f^{\beta}$, where the different values of $\beta$ are associated with ingoing and
outgoing wave solutions. Near the AdS boundary $u=0$, solutions that satisfy the ingoing-wave condition
at the horizon behave as
\begin{equation}
Z_{p}(u) = \mathcal{A} (1 + \cdots) u^{\alpha_{1}} + \mathcal{B} (1 + \cdots) u^{\alpha_{2}},
\end{equation}
where the exponents $\alpha_1$ and $\alpha_2$ correspond, respectively, to the leading terms of the non-normalizable and the normalizable solutions.

The metric and scalar field perturbations take the same form as defined in Section \ref{Sec:gaugeInvQuantities},
and the process of finding the master equation in each sector is identical. Furthermore, the equations of motion
for the background are given by \eqref{Eq:MotionZero}--\eqref{Eq:Phiorderzero}, whose solution has already
been determined.

\section{The tensor sector of perturbations}\label{Subsec:TThydro}

\subsection{Master equation and hydrodynamic limit}

The tensor sector is associated with the transverse-traceless part of the metric perturbations. 
In the present case, there is only one independent perturbation of this kind, namely, $H_{12}$.
In this sector, the perturbation function $H_{12}$
is a gauge-invariant variable (see Appendix \ref{app:Invariant} for details).
By substituting $\zeta_1=\zeta_2$ and $Z_T = H_{12}$ in Eq.~\eqref{Eq:Hxy},
the master equation becomes
\noindent
\begin{align} \label{Eq:ZTmaster}
    Z_{T}''&+\left(\frac{f'}{f}-\frac{f'' }{f'}\right) Z_{T}'+\frac{\mathfrak{w}^2-\mathfrak{q}^2 f}{ f^2}Z_{T}=0,
    \end{align}
\noindent
which decouples naturally from the other equations in the system. 

Considering a change of the independent variable of the general form \eqref{Eq:GeneralSolutionZ} and imposing
on $Z_{T}(u)$ the infalling-wave condition at the horizon, we have $Z_{T}(u)=f(u)^{i\mathfrak{w}/f'_{h}} F_{T}(u)$.
Here, $F_{T}(u)$ is a regular function at the horizon that satisfies the following second-order differential equation: 
\begin{equation}
    \begin{aligned}
          F_{T}'' - \bigg[\frac{f''}{f'} - & \frac{f'\left(f'_{h} - 2 i \mathfrak{w}\right)}{ff'_{h}}\bigg] F_{T}'\\
          &+\bigg[\frac{\mathfrak{w}^2}{f^2} \bigg(1-\frac{{f'}^2}{{f'_{h}}^{2}}\bigg)-\frac{\mathfrak{q}^2}{f}\bigg]F_{T}=0.
    \end{aligned}
\label{Eq:FTTu}
\end{equation}

To study the hydrodynamic limit of the tensor perturbations, we propose constructing a solution to
Eq.~\eqref{Eq:FTTu} as a power series in the frequency $\mathfrak{w}$ and wavenumber $\mathfrak{q}$. Specifically,
we reparameterize the differential equation \eqref{Eq:FTTu} by introducing the substitutions
$\mathfrak{w}\to\lambda\mathfrak{w}$ and $\mathfrak{q}\to \lambda \mathfrak{q}$, where $\lambda$ is a small parameter
($\lambda \ll 1$). We then posit an ansatz for the solution of the form
\begin{equation}
F_{T}(u)=\sum_{n=0}^{\infty} \lambda^n F_{T}^{n}(u),
\end{equation}
where the coefficient functions $F_T^n(u)$ (for $n=0,\, 1,\,\dots$) are to be determined.
The parameter $\lambda$ facilitates organizing the multiparameter expansion of $F_{T} (u)$ in powers of
$\mathfrak{w}$ and $\mathfrak{q}$.

At zeroth order in $\lambda$, the resulting equation for $F_T^0(u)$ is
\begin{gather}
    F^{0}_{T}{}''+ \left(\frac{f'}{f}-\frac{f''}{f'}\right)F^{0}_{T}{}'=0,
    \label{Eq:FTT0}
    \end{gather}
and the general solution to this equation is
\begin{gather}
    F^{0}_{T}(u)=c_1+c_2\log f(u),
    \label{Eq:solutionTT0}
\end{gather}
where $c_1$ and $c_2$ are integration constants.
Applying the condition of regularity at the horizon ($u=1$) to the solution \eqref{Eq:solutionTT0} requires $c_2=0$,
since $f(1)=0$ and $\log f(u)$ diverge as $u\rightarrow 1$. Therefore, at the zeroth order in $\lambda$,
the regular solution is simply a constant: $F^{0}_{T}(u)=c_1$.

At first order in $\lambda$, Eq.~\eqref{Eq:FTTu} yields 
\begin{equation}
    F^{1}_{T}{}''+\left(\frac{f'}{f}-\frac{f''}{f'}\right)F^{1}_{T}{}'-\frac{2 i \mathfrak{w} f'F^{0}_{T}{}'}{ff'_{h}}=0.
    \label{Eq:FTT1}
\end{equation}
Now, considering that $F^{0}_{T}$ is a constant, the last term in \eqref{Eq:FTT1} vanishes. With this,
the differential equation for $F^1_T(u)$ becomes identical to Eq.~\eqref{Eq:FTT0}, which means that $F_T^1$
is also a constant. Thus, the solution for $Z_{T}(u)$ up to first order in the expansion parameter $\lambda$
(which tracks the powers of $\mathfrak{w}$ and $\mathfrak{q}$) may be written as 
\begin{equation}
    Z_{T}(u)=\mathcal{C}_{T}f^{i\mathfrak{w}/f'_{h}}\left[1+\mathcal{O}\left(\mathfrak{q}^{2},\mathfrak{w}^{2},\mathfrak{w}\mathfrak{q}\right)\right],
\end{equation}
where $\mathcal{C}_{T}$ is an integration constant. 

Applying the Dirichlet condition $Z_{T}=0$ at the boundary ($u=0$),
we identify the coefficients as $\mathcal{A}=1$ and $\mathcal{B}=(i \mathfrak{w}/2 f'_{h})\left[f'^2(0)-f''(0)\right]=0$.
This indicates that the tensor sector does not present hydrodynamic modes, as expected for transverse-traceless perturbations.

\begin{table*}
\centering
\begin{tabular}{l |c|c|c|c|c}
\hline
\hline
$n$ & $\mathfrak{q}=0$, $\phi_{h}=0$ & $\mathfrak{q}=1$, $\phi_{h}=0$ & $\mathfrak{q}=2$, $\phi_{h}=0$ & $\mathfrak{q}=3$, $\phi_{h}=0$ & $\mathfrak{q}=4$, $\phi_{h}=0$ \\
\hline 
$1$ & $\pm1.55973 - 1.37334i $  & $\pm1.67202 - 1.34041i$ & $\pm1.95433 - 1.26733i$ & $\pm 2.32581-1.18941 i$ & $\pm 2.74191 -1.12045 i$  \\
$2$ & $\pm2.58476 - 2.38178i $  & $\pm2.66585 - 2.35892i$ & $\pm2.88026 - 2.29796i$ & $\pm 3.17957-2.21743 i$ & $\pm 3.53159 -2.13311 i$\\
$3$ & $\pm3.59362 - 3.38604i $  & $\pm3.65954 - 3.36769i$ & $\pm3.83724 - 3.31526i$ & $\pm 4.09071-3.24014 i$ & $\pm 4.39653 -3.15479 i$  \\
$4$ & $\pm4.70155 - 4.27078i $  & $\pm4.72988 - 4.25696i$ & $\pm4.82127 - 4.22576i$ & $\pm 5.02766-4.25856 i$ & $\pm 5.30261 -4.17723 i$ \\
\hline 
\hline
  $n$ & $\mathfrak{q}=0$, $\phi_{h}=0.4$ & $\mathfrak{q}=1$, $\phi_{h}=0.4$ & $\mathfrak{q}=2$, $\phi_{h}=0.4$ & $\mathfrak{q}=3$, $\phi_{h}=0.4$ & $\mathfrak{q}=4$, $\phi_{h}=0.4$ \\
\hline 
 $1$ & $\pm1.5541 - 1.30195i$  & $\pm1.66367 - 1.28031i$ & $\pm1.94402 -1.22699i$ & $\pm 2.31672-1.16308i$ & $\pm 2.7347 -1.10188 i$\\
 $2$ & $\pm2.58938 - 2.2353i$  & $\pm2.66272 - 2.22641i$ & $\pm2.86634 -2.19656i$ & $\pm 3.16269-2.14536 i$ & $\pm 3.51662 -2.08103 i$  \\
 $3$ & $\pm3.61616 - 3.16194i$  & $\pm3.67063 - 3.15759i$ & $\pm3.82803 -3.14113i$ & $\pm 4.07056-3.1068 i$ & $\pm 4.3746 -3.05433 i$  \\
 $4$ & $\pm4.65264 - 4.08445i$  & $\pm4.69412 - 4.08005i$ & $\pm4.81519 -4.06733i$ & $\pm 5.01236-4.05328 i$ & $\pm 5.27635 -4.01601i$ \\
\hline
\hline
$n$ & $\mathfrak{q}=0$, $\phi_{h}=1.0$ & $\mathfrak{q}=1$, $\phi_{h}=1.0$ & $\mathfrak{q}=2$, $\phi_{h}=1.0$ & $\mathfrak{q}=3$, $\phi_{h}=1.0$ & $\mathfrak{q}=4$, $\phi_{h}=1.0$ \\
\hline 
 $1$ & $\pm 1.61890 -  1.03961 i$  & $\pm 1.70802 - 1.03819 i$ & $\pm 1.95157-1.03203 i$ & $\pm 2.30012-1.01772 i$ & $\pm 2.70946 -0.994557 i$  \\
 $2$ & $\pm 2.76603 -1.83537 i$  & $\pm 2.81830-1.83340 i$ & $\pm 2.97043-1.82823 i$ & $\pm 3.21034-1.82047i$ & $\pm3.521960 -1.80855 i$  \\
 $3$ & $\pm 3.88202 - 2.64418 i$  & $\pm 3.92017-2.64186 i$ & $\pm 4.03263-2.63547 i$ & $\pm 4.21409-2.62620 i$ & $\pm 4.45731 -2.61471 i$  \\
 $4$ & $\pm 4.98390 - 3.45533 i$  & $\pm 5.01434-3.45314 i$ & $\pm 5.10458-3.44689 i$ & $\pm 5.25164-3.43736 i$ & $\pm 5.45129 -3.42536 i$  \\
\hline
\end{tabular}
\caption{
The quasinormal frequencies of the tensor sector for a few selected values of the wavenumber and for three values of the dilaton strength
$\phi_{h}$, namely, $\phi_{h}=0$ (top), $\phi_{h}=0.4$ (center), and $\phi_{h}=1.0$ (bottom). The results for $\phi_{h}=0$ and $\mathfrak{q}=2$
agree with those of Ref.~\cite{Kovtun:2005ev}.}
\label{Tab:QNMsTT}
\end{table*}

\subsection{Numerical results for the QNMs}

Here we present and analyze the results obtained by solving the eigenvalue problem for the transverse and traceless sector.
As mentioned previously, to apply the numerical methods of Ref.~\cite{Jansen:2017oag}, the master equation must be expressed
in terms of the advanced E-F time coordinate. This transformation is equivalent to replacing the original perturbation function
$Z_T(u)$ with a new function $\hat{Z}_T(u)=\exp[-i\mathfrak{w}u^*]Z_T(u)$. Therefore, the master equation
satisfied by the transformed function $\hat{Z}_T(u)$ is
\begin{equation}
    \hat{Z}_{T}''+ \left(\frac{f'}{f}-\frac{f''}{f'}+\frac{2 i \mathfrak{w}}{f} \right)\hat{Z}_{T}'
    - \left(\frac{\mathfrak{q}^2}{f}+\frac{i \mathfrak{w}  f''}{ff'}\right)\hat{Z}_{T}=0. 
\end{equation}

The new function $\hat{Z}_T(u)$ must satisfy the usual boundary conditions: ingoing waves at the horizon and
appropriate falloff at the AdS boundary. We first analyze the regularity condition at the horizon ($u=1$)
by examining solutions of the form $\hat{Z}_{T}  \sim f^{\beta}$. The characteristic exponents are found to be
\begin{equation}
    \beta_{1}=0 \, , \qquad \beta_{2}=-\frac{2 i \mathfrak{w}}{f'_{h}}.
\end{equation}
Both values of $\beta$ correspond to functions that are regular at the horizon. However, the solution associated with $\beta_{1}$
represents purely infalling waves at the horizon, which is the physically required condition for quasinormal modes, while the solution for $\beta_{2}$ includes outgoing waves. Therefore, the appropriate solutions for $\hat{Z}_T(u)$ near the horizon behave as
$\hat{Z}_T(u)\sim f(u)^0 =\text{constant}$. 

Near the AdS boundary, we propose a solution of the form $\hat{Z}_{T}\sim u^{\alpha}$ and consider $f(u) \to 1-u^{4}$ as $u\to 0$.
The analysis yields characteristic exponents $0$ and $4$, where $\alpha=0$ is associated to a non-normalizable
mode, while $\alpha=4$ corresponds to a normalizable mode, as it vanishes at the boundary $u = 0$, according to
the general argument presented in Section \ref{Sec:QNMgeneral}. In addition, to improve numerical stability, following
Ref.~\cite{Jansen:2017oag}, we can rescale the variable as $\tilde{Z}_{T} = \hat{Z}_{T}/u$. Applying the same near-boundary ansatz $\tilde{Z}_{T}=u^{\tilde{\alpha}}$ to the equation for $\tilde{Z}_{T}$ leads to the exponents $\tilde{\alpha}=-1$
and $\tilde{\alpha}=3$. This alternative analysis confirms that the mode corresponding to the original non-normalizable solution
($\hat{Z}_T \sim u^0$) behaves as $\tilde{Z}_T \sim u^{-1}$ in the rescaled variable, while the mode corresponding to the original
normalizable solution ($\hat{Z}_T \sim u^4$) behaves as $\tilde{Z}_T \sim u^{3}$. This rescaling preserves the boundary conditions
and improves the numerical stability, while the numerical results remain unchanged \cite{Jansen:2017oag}.

\begin{figure} [ht]
    \centering
    \includegraphics[width=7cm]{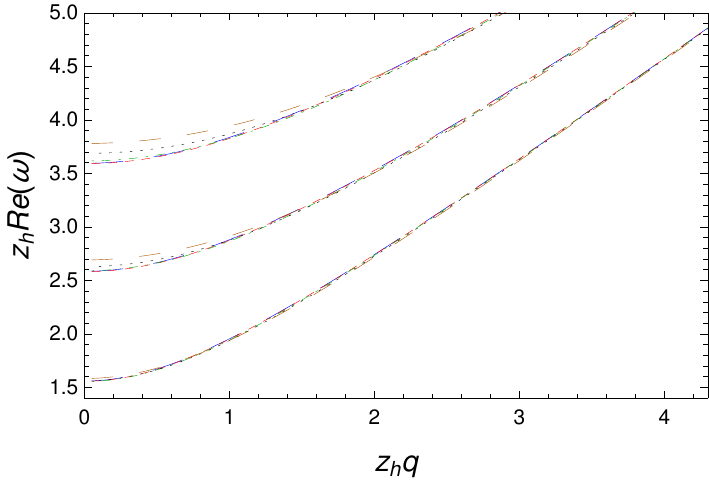}
    \includegraphics[width=7cm]{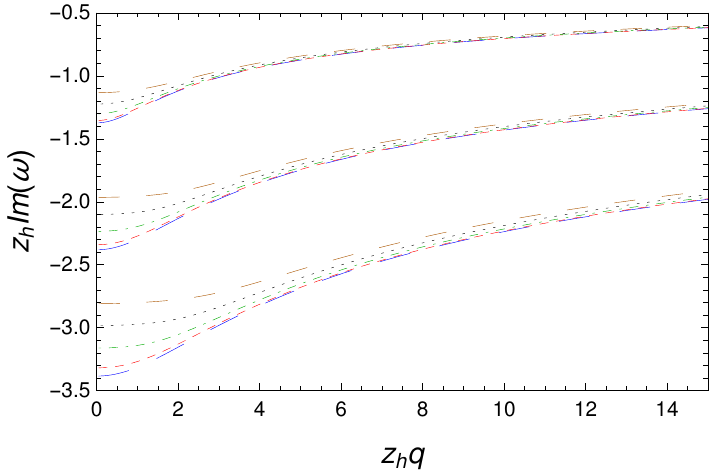}
    \caption{The real and imaginary parts of the first three frequencies of the
    quasinormal
    modes of the tensor sector as a function of wavenumber $\mathfrak{q} = z_{h}q$, for $\phi_{h}=0$ (blue long-dashed lines),
    $\phi_{h}=0.2$ (red dashed lines), $\phi_{h}=0.4$ (green dot-dashed lines), $\phi_{h}=0.6$ (black dotted lines),
    and $\phi_{h}=0.8$ (dark orange long-dashed lines).}
    \label{Fig:QNMTTomegaVsq}    
\end{figure}

Our numerical results for the QNM frequencies of the tensor sector are shown in Figure \ref{Fig:QNMTTomegaVsq} and
Table \ref{Tab:QNMsTT}. As usual, the frequency $\mathfrak{w}$ is decomposed into real and imaginary parts as
$\mathfrak{w}=z_{h}\left[\text{Re}(\omega)-i \text{Im}(\omega)\right]$.

In Figure \ref{Fig:QNMTTomegaVsq} we present the real (top panel) and imaginary (bottom panel) parts of $\mathfrak{w}=z_{h} \omega$
as a function of $\mathfrak{q} = z_{h}q$. The first three modes are shown for selected values of $\phi_{h}$ in the interval $[0, 1]$. We observe that higher modes are more strongly affected by the dilaton, particularly for small values of $\mathfrak{q}$. This effect is more evident
in the imaginary part of $\mathfrak{w}$, which becomes less negative as $\phi_{h}$ increases. The larger modification in the imaginary
part of $\mathfrak{w}$ implies that the dilaton has a greater influence at small $\mathfrak{q}$. As $\mathfrak{q}$ increases,
the influence of the dilaton decreases, and the frequencies tend to converge to the conformal value (corresponding to $\phi_{h}=0$).

In Table \ref{Tab:QNMsTT} we show the results for the first four quasinormal modes for five selected values of $\mathfrak{q}$, namely $\mathfrak{q}=0,\,1,\,2,\,3,\,4$,
and three selected values of $\phi_h$, namely $\phi_{h}=0,\, 0.4,\, 1.0$. The case with $\mathfrak{q}=2$ and $\phi_{h}=0$ is consistent with the results
obtained in Ref.~\cite{Kovtun:2005ev} (K-S) in the conformal limit.

\section{The vector sector of perturbations}\label{Subsec:Vecthydro}

\subsection{Master equation and hydrodynamic limit}

Perturbations associated with the vector sector provide information about how the fluid behaves under shear stress,
and the shear viscosity can be extracted from the behavior of these perturbations in the hydrodynamic limit. 
The master equation is obtained using the gauge-invariant quantity defined in Eq.~\eqref{deltav}, and the
simplification process is schematically explained in Appendix \ref{Ap:vector}. We obtain 
\begin{gather}\label{Eq:masterVector}
    Z_{V}''+\left[\frac{\mathfrak{w}^2 f'}{f\left(\mathfrak{w} ^2-\mathfrak{q}^2 f\right)}
    -\frac{f''}{f'}\right]Z_{V}'+\frac{ \left(\mathfrak{w}^2-\mathfrak{q}^2 f\right)}{f^2 }Z_{V}=0. 
\end{gather}
Assuming the general solution form presented in Eq.~\eqref{Eq:GeneralSolutionZ}, $Z_{V}(u)=f(u)^{i\mathfrak{w}/f'_{h}} F_{V}(u)$,
where $F_{V}(u)$ must be a regular function at the horizon, the resulting differential equation for $F_{V}(u)$ is
\begin{equation} 
    \begin{aligned}
        F_{V}'' & + \left[\frac{2i\mathfrak{w} f'}{f'_{h} f}+\frac{\mathfrak{w}^2 f'}{f\left(\mathfrak{w}^2-\mathfrak{q}^2 f\right)}-\frac{f''}{f'}\right] F_{V}'
        +\bigg[\frac{\left(\mathfrak{w}^2-\mathfrak{q}^2 f\right)}{f'^2}\\
	&+\frac{i\mathfrak{w}^3}{f'_{h}\left(\mathfrak{w}^2-\mathfrak{q}^2 f\right)}+\frac{i\mathfrak{w}}{f'_{h}}\left(\frac{i\mathfrak{w}}{f'_{h}}-1\right)\bigg]
        \left(\frac{f'}{f}\right)^2 F_V =0.
    \end{aligned}
\end{equation}

Following a similar procedure as for the tensor sector presented in the previous section,
we propose a solution of the form $F_{V}(u)=\sum_{n=0}^{\infty} \lambda^n F_{V}^{n}(u)$. 
Thus, the differential equation at the zeroth order in $\lambda$ is
\begin{equation}
    F^{0}_{V}{}''+ \left[\frac{\mathfrak{w}^2 f'}{f\left(\mathfrak{w}^2-\mathfrak{q}^2 f\right)} - \frac{f''}{f'}\right]F^{0}_{V}{}'=0,
\end{equation}
whose general solution is 
\begin{equation}\label{Eq:FV0}
    F^{0}_{V}=c_{3}-c_{4} \mathfrak{q}^2 f + c_{4} \mathfrak{w}^2 \log f,
\end{equation}
where $c_3$ and $c_4$ are integration constants. The regularity condition at the horizon ($u=1$)
requires that $c_4=0$. Therefore, the regular zeroth-order solution is simply a constant: $F^{0}_{V}=c_3$.

At first order in $\lambda$, the resulting differential equation for $F^{1}_{V}$ reads
\begin{equation}\label{Eq:FV1}
\begin{aligned}
     F^{1}_{V}{}''+ & \bigg[\frac{\mathfrak{w}^2 f'}{f\left(\mathfrak{w}^2 -\mathfrak{q}^2 f\right)}-\frac{f''}{f'}\bigg]F^{1}_{V}{}' \\
     & +\frac{i \mathfrak{w}\mathfrak{q}^2 f'^2}{f'_{h}f(\mathfrak{w}^{2}-\mathfrak{q}^{2}f)}F^{0}_{V}(u)+\frac{2 i \mathfrak{w} f'}{f'_{h}f} F^{0}_{V}{}'=0.
\end{aligned}
\end{equation} 
Now, taking into account that $F^{0}_{V}$ is a constant, the last term in \eqref{Eq:FV1} vanishes, and the general solution is 
\begin{equation}
    F^{1}_{V}{}=c_{5}-c_{6} \left(\mathfrak{q}^2 f - \mathfrak{w}^2 \log f\right) -\frac{i \mathcal{C}_V \mathfrak{w} \log f}{f'_{h}}, 
\end{equation}
where $c_5$ and $c_6$ are integration constants, and we have renamed the constant $c_3$ to $\mathcal{C}_V$.  
The condition of regularity of $Z_V$ at the horizon requires $c_5=0$ and leads to $c_{6}= i\mathcal{C}_{V}/(\mathfrak{w} f'_{h})$.
Then, the hydrodynamic limit of the ingoing-wave solution at the horizon is given by
\begin{equation}\label{Eq:Z1f}
    Z_{V}=\mathcal{C}_{V}f^{i\mathfrak{w}/f'_{h}} \left(1-\frac{i \mathfrak{q}^2 f}{\mathfrak{w} f'_{h}}\right).
\end{equation}

As mentioned previously, we want to identify the coefficient $\mathcal{A}(\mathfrak{w},\mathfrak{q})$ of the non-normalizable
mode at the boundary ($u\rightarrow 0$). The zeros of $\mathcal{A}$
correspond to poles in the retarded two-point function of the associated boundary operator. For this,
we expand \eqref{Eq:Z1f} near $u=0$ and consider the limit $\mathfrak{w},\mathfrak{q}\rightarrow 0$ to get
$\mathcal{A}(\mathfrak{w},\mathfrak{q})=\mathcal{C}_V [1-i\mathfrak{q}^2/(\mathfrak{w}f'_{h})]$.
The condition $\mathcal{A}(\mathfrak{w},\mathfrak{q})=0$ leads to 
\begin{gather}\label{Eq:Shearuniversal}
\mathfrak{w}=\frac{i \mathfrak{q}^{2}}{f'_{h}}, \qquad \text{or} \qquad
\omega=-i\frac{q^2}{4\pi T}\equiv -i\gamma_{\eta} q^2.
\end{gather}
This solution is compatible with the hydrodynamic shear mode. From the dispersion relation \eqref{Eq:Shearuniversal},
we can identify the diffusion coefficient $\gamma_\eta \equiv \eta/(s T)=1/(4\pi T)$ and, as expected for this sector,
we obtain the universal relation $\eta/s=1/(4\pi)$.

One way to estimate the contribution of the dilaton to the dispersion relation \eqref{Eq:Shearuniversal}, 
is by analyzing the behavior of the Hawking temperature in the limit $\phi_{h} \rightarrow 0$. In such a limit,
the horizon function can be approximated by
\begin{equation}
f(u)=\left(1-u^4\right)+\frac{2u^{4}}{15}\left(1- u^4\right) \phi_{h}^2,
\end{equation}
whose derivative evaluated at $u=1$ is
\begin{equation}
f'_{h}=-\left(4+\frac{8}{15} \phi_{h}^2\right).
\end{equation}
Then, substituting this into the dispersion relation yields,
\begin{equation}\label{Eq:shearrelationwithc}
   \mathfrak{w}=-i\gamma_{\eta}\mathfrak{q}^{2}=-i\left(\frac{1}{4}-\frac{1}{30}  \phi_{h}^2\right) \mathfrak{q}^{2}, 
\end{equation}    
that shows explicitly the modification of the diffusion coefficient $\gamma_{\eta}$ caused by a non-vanishing dilaton field.

Starting from the relation $\eta/s=1/(4\pi)$ and using Eqs. \eqref{Eq:Entropy} and \eqref{eq:temp-zh},
we obtain the following expression for the normalized shear viscosity:
\begin{equation}
\frac{\eta}{N_{c}^{2}T^{3}}=\frac{M_{p}^{3}C_h(T)}{4\pi T^4},
\end{equation}
where $C_h=C_h(\phi_h(T))$ is a function of the temperature,
defined by Eq.~\eqref{Eq:Ch}.

\begin{figure}[h!]
\centering
\includegraphics[width=7cm]{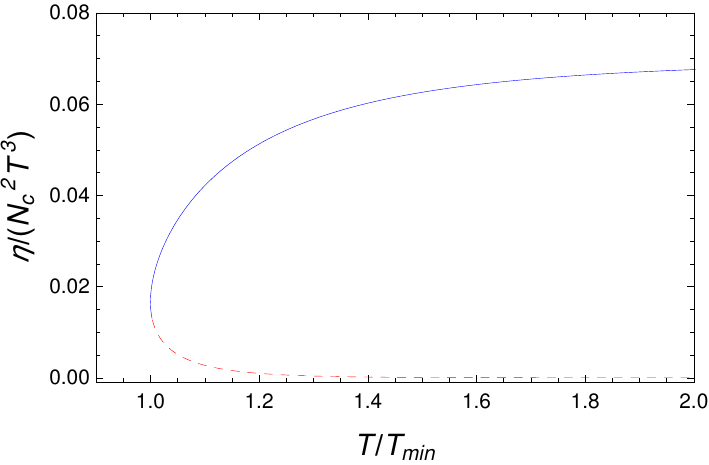}
\caption{The shear viscosity normalized by $N_{c}^{2}T^{3}$ as a function of the normalized temperature.
Large (Small) black hole solutions correspond to the solid (dashed) blue (red) line.}
\label{Fig:ShearViscosity}
\end{figure}

In Figure \ref{Fig:ShearViscosity} shows the shear viscosity $\eta$, normalized by $N_{c}^{2}T^{3}$, as a function of the
temperature normalized by its minimum value, $T/T_{min}$, where we set $M_p^3=1/(45\pi^2)$ \cite{Gursoy:2008za}. 
The solid blue line represents the large black hole regime,
whereas the dashed red line represents the small black hole regime. The overall shape of this
curve exhibits a similar dependence to that obtained in Ref.~\cite{Ballon-Bayona:2021tzw}.

\subsection{Numerical results for the QNMs}

Similarly to the tensor sector, to calculate the QNM frequencies of the vector perturbation sector, it is convenient to rewrite the master differential equation in terms of the advanced E-F time, which is equivalent to replacing the function $Z_V(u)$ in Eq.~\eqref{Eq:masterVector} by
$\hat{Z}_V(u)=\exp[-i\mathfrak{w}u^*]Z_V(u)$

Therefore, the master equation satisfied by the transformed function $\hat{Z}_V(u)$ is
\begin{equation}\label{Eq:ZetaVectorQNM}
    \begin{aligned}
       \hat{Z}_{V}''-\bigg[ &\frac{f''}{f'} -\frac{2i\mathfrak{w}}{f} +  \frac{ \mathfrak{w} ^2 {f'}}{f\left(\mathfrak{q}^2 f-\mathfrak{w}^2\right)}\bigg] \hat{Z}_{V}'\\
        &-\left[\frac{i\mathfrak{w}f''}{f f'} + \frac{q^2}{f} +\frac{i \mathfrak{q}^2 \mathfrak{w}  {f'}}{f\left(\mathfrak{q}^2 f-\mathfrak{w} ^2\right)}\right] \hat{Z}_{V}=0. 
    \end{aligned}
\end{equation}
Compared to the original master equation \eqref{Eq:masterVector}, the main difference in Eq.~\eqref{Eq:ZetaVectorQNM} is the appearance
of terms explicitly linear in $\mathfrak{w}$ in the coefficients of $Z'_V$ and $Z_V$, arising from the change to E-F time coordinate.

The behavior of $\hat{Z}_V$ near the event horizon ($f\rightarrow 0$) is analyzed using the ansatz $\hat{Z}_{V}\sim f^{\beta}$,
where $\beta$ takes the same values as for the tensor sector, $\beta_{1}=0$ and $\beta_{2}=-2 i \mathfrak{w}/f'_{h}$.
The exponent $\beta_{1}=0$ corresponds to the physically relevant ingoing-wave condition at the horizon required for QNMs,
while $\beta_{2}=-2 i \mathfrak{w}/f'_{h}$ corresponds to outgoing waves.

Near the AdS boundary ($u\rightarrow 0$), we look for a solution of the form $\hat{Z}_{V}=u^{\alpha}$.
In this limit, the blackening function behaves as $f(u)\sim 1-u^{4}$.
The analysis yields the characteristic exponents $\alpha_1=0$ and $\alpha_2=4$. The solution with $\alpha_1=0$ corresponds to the non-normalizable mode and is neglected, while the solution with $\alpha_2=4$ corresponds to the normalizable mode.
As discussed for the tensor sector, rescaling the variable as $\tilde{Z}_{V} = \hat{Z}_{V}/u$ for improved numerical stability
leads to corresponding exponents $\tilde{\alpha}= -1$ (non-normalizable) and $\tilde{\alpha}= 3$ (normalizable), consistent with the original analysis.

The quasinormal frequencies were calculated using two different grid sizes to verify the consistency of the results, and the calculations were
performed for several values of the dilaton parameter $\phi_{h}$ and the wavenumber $\mathfrak{q}$.

\begin{figure}[h!]
        \centering
        \includegraphics[width=7cm]{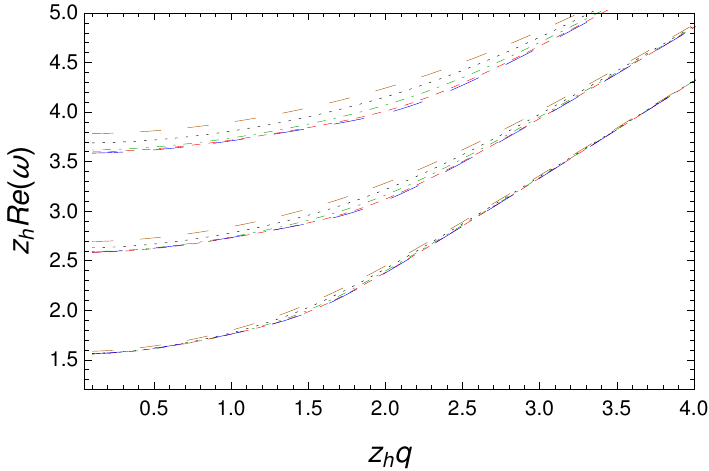}
        \label{Fig:QNMvectorREomegaVsq} 
        \includegraphics[width=7cm]{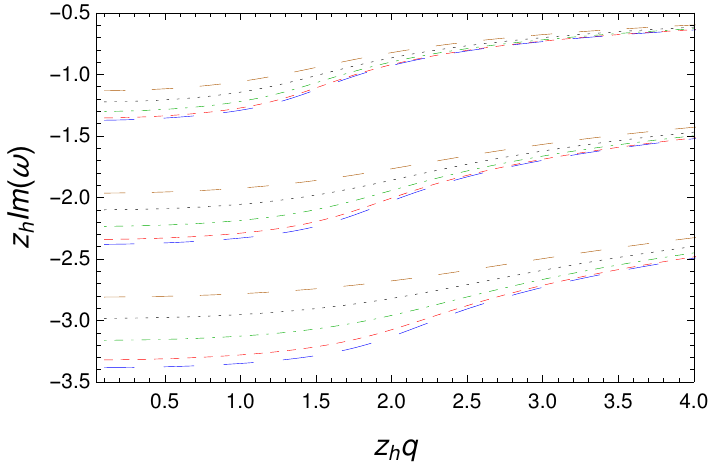}
    \caption{The real (top panel) and imaginary (bottom panel) parts of the frequencies $\mathfrak{w}=z_h \omega$
    of the vector sector as a function of wavenumber $\mathfrak{q} = z_{h}q$. The figure shows the frequencies of
    the first three non-hydrodynamic modes for five different values of $\phi_{h}$, namely, $\phi_{h}=0$
    (blue long-dashed), $\phi_{h}=0.2$ (red dashed lines), $\phi_{h}=0.4$ (green dot-dashed lines), $\phi_{h}=0.6$
    (black dotted lines), and $\phi_{h}=0.8$ (dark orange long-dashed lines).}   
    \label{Fig:QNMvectorImomegaVsq}
\end{figure}

The numerical results for the non-hydrodynamic QNM frequencies of the vector perturbations are displayed in
Fig.~\ref{Fig:QNMvectorImomegaVsq}.
This figure displays the real (top panel) and imaginary (bottom panel) parts of the frequency $\mathfrak{w}$ for the first three non-hydrodynamic modes
as a function of $\mathfrak{q}$ for selected values of $\phi_h$. The real part generally increases with $\mathfrak{q}$, showing relatively small variations with $\phi_h$. In contrast, the imaginary part (representing the decay rate) is more sensitive to $\phi_h$, especially at small $\mathfrak{q}$, where larger $\phi_h$ leads to less negative values (slower decay).
For large $\mathfrak{q}$, the influence of the dilaton diminishes, and the imaginary parts appear to approach a common asymptotic behavior.

\begin{figure}[h!]
 \centering
 \includegraphics[width=7cm]{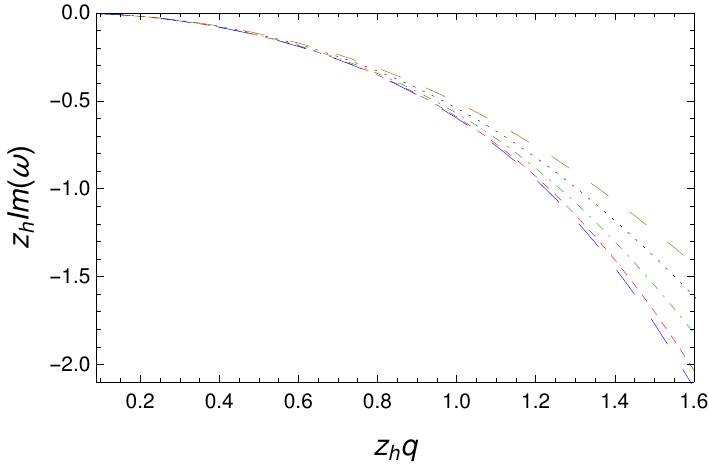}
 \caption{The imaginary part of the frequency of the hydrodynamic QNM for five different values of $\phi_{h}$,
 namely, $\phi_{h}=0$ (blue long-dashed), $\phi_{h}=0.2$ (red dashed lines), $\phi_{h}=0.4$ (green dot-dashed lines),
 $\phi_{h}=0.6$ (black dotted lines), and $\phi_{h}=0.8$ (dark orange long-dashed lines).}
 \label{Fig:HydroMode}
\end{figure}

\begin{figure}[h]
 \centering
 \includegraphics[width=7cm]{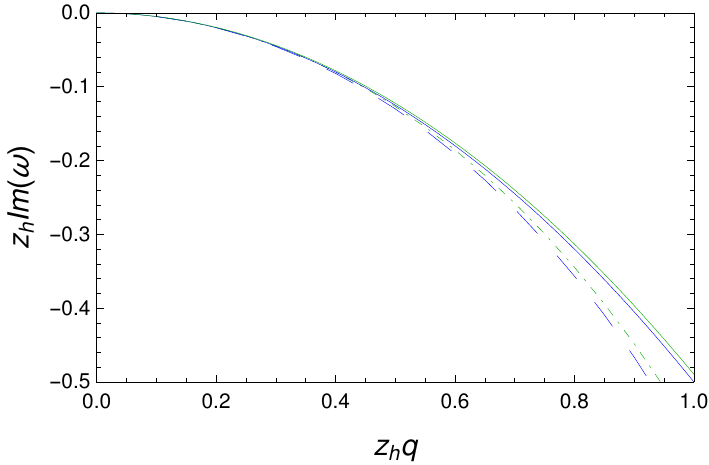}
 \caption{The imaginary part of the frequency of the hydrodynamic QNM obtained numerically (dashed lines) compared to the approximate result
 \eqref{Eq:Shearuniversal} (solid lines) for $\phi_{h}=0$ (blue lines) and  $\phi_{h}=0.4$ (green lines)}
 \label{Fig:ModoZeroShear}
\end{figure}

Regarding the hydrodynamic QNM, the real part of the frequency remains zero, independently of the dilaton strength,
as expected from the dispersion relation for this sector. On the other hand, the imaginary part, displayed in Figure
\ref{Fig:HydroMode}, shows a different dependence on $\mathfrak{q}$ and $\phi_h$ compared to
the non-hydrodynamic modes: the influence of the dilaton parameter appears more pronounced at larger
values of $\mathfrak{q}$, while the results are less sensitive to $\phi_h$ at small $\mathfrak{q}$.

Figure \ref{Fig:ModoZeroShear} compares the numerically calculated hydrodynamic QNM frequencies (dashed lines) with the analytical
approximation derived from Eq.~\eqref{Eq:Shearuniversal}, represented by solid lines, for $\phi_h=0$ and $\phi_h= 0.4$.
The figure shows excellent agreement for small $\mathfrak{q}$, while for larger $\mathfrak{q}$, the difference between the analytical approximation and the numerical results becomes apparent. This deviation at larger $\mathfrak{q}$ is expected, as the analytical formula is derived under strict hydrodynamic limit assumptions ($\mathfrak{w}\ll 1$,$\mathfrak{q} \ll 1$), which may not hold precisely as $\mathfrak{q}$ increases.

\begin{table*}
\centering
\begin{tabular}{l|c|c|c|c}
\hline
\hline
$n$ & $\mathfrak{q}=1$, $\phi_{h}=0$ & $\mathfrak{q}=2$, $\phi_{h}=0$ & $\mathfrak{q}=3$, $\phi_{h}=0$ & $\mathfrak{q}=4$, $\phi_{h}=0$ \\
\hline 
$1$ & $\pm0 -0.13007i$ & $\pm0 -0.59806 i$ & $\pm 0 -1.76691i$ & $\pm 0-4.21761 
i$ \\
$2$ & $\pm1.61638 -1.35509i$ & $\pm1.75912 -1.29159 i$ & $\pm 2.87557-2.23602i$ & $\pm 4.2819 -4.27324i$ \\
$3$ & $\pm2.62558 -2.36986i$ & $\pm2.73308 -2.33041 i$ & $\pm 3.83301 -3.28577i$ & $\pm 5.84507 -4.37374 i$ \\
$4$ & $\pm 3.61718 -3.37689i$ & $\pm 3.71587 -3.34501 i$ & $\pm 4.93923-3.98277i$ & $\pm 6.69348 -4.60543i$ \\
\hline 
\hline
  $n$ &  $\mathfrak{q}=1$, $\phi_{h}=0.4$ & $\mathfrak{q}=2$, $\phi_{h}=0.4$ & $\mathfrak{q}=3$, $\phi_{h}=0.4$ & $\mathfrak{q}=4$, $\phi_{h}=0.4$ \\
\hline 
 $1$ & $\pm0-0.12682i$ & $\pm0-0.570549i$ & $\pm 0 -1.54927i$ & $\pm 0-3.11817i$\\
 $2$ & $\pm1.61036-1.28429i$ & $\pm1.75859 -1.21994i$ & $\pm 2.90730-2.09905i$ & $\pm 5.08193 -3.85816i$ \\
 $3$ & $\pm2.62926-2.22458i$ & $\pm2.73985 -2.18659i$ & $\pm 3.87717-3.06895i$ & $\pm 5.88371 -4.19243i$ \\
 $4$ & $\pm3.64675-3.15392i$ & $\pm3.73824 -3.12674i$ & $\pm 4.94850-3.93540i$ & $\pm 6.75506 -4.35293i$ \\
\hline\hline
\end{tabular}
\caption{
The quasinormal frequencies of the vector sector for selected values of the wavenumber $\mathfrak{q}$ for $\phi_{h}=0$ (top) and $\phi_{h}=0.4$ (bottom).
The results for $\phi_{h}=0$ agree with those of Ref.~\cite{Kovtun:2005ev}. 
}
\label{Tab:QNMsVectorial}
\end{table*}

\begin{table*} 
\centering
\begin{tabular}{c|c|c|c|c|c}
\hline\hline
$\mathfrak{q}$ & Model ($\phi_{h}=0$) & Model ($\phi_{h}=0.2$) & Model ($\phi_{h}=0.4$) & Model ($c\phi_{h}=0.6$) & Model ($\phi_{h}=0.8$)\\
\hline
$1$ & $-0.13007i$ & $-0.12924i$ & $-0.12682i$ & $-0.12299i$ & $-0.11803i$  \\
\hline
$2$ & $-0.59806i$ & $-0.59075i$ & $-0.57055i$ & $-0.54152i$ & $-0.50785i$ \\
\hline
$3$ & $-1.76691i$ & $-1.69912i$ & $-1.54927i$ & $-1.38790i$ & $-1.24229i$  \\
\hline
$4$ & $-4.21761i$ & $-3.72279i$ & $-3.11817i$ & $-2.63571i$ & $-2.27545i$ \\
\hline\hline
\end{tabular}
\caption{The hydrodynamic QNM frequencies of the vector sector for different values of the wavenumber $\mathfrak{q}$ and several different
values of the dilaton constant $\phi_{h}$. Note that the results for $\phi_{h}=0$ and $\mathfrak{q}=2$ are in agreement with
the results of Ref.~\cite{Kovtun:2005ev}.
}
\label{Tab:HydroQNMVector}
\end{table*}

Table \ref{Tab:QNMsVectorial} displays numerical results for the first four modes ($n=1,\,2,\,3,\,4$) for some selected values of $\mathfrak{q}$ and $\phi_h$, namely $\mathfrak{q}=1,\,2,\,3,\,4$ and $\phi_{h}=0,\,0.4$.
It is worth mentioning that the results for $\mathfrak{q}=2$ and $\phi_{h}=0$ agree with those reported in Ref.~\cite{Kovtun:2005ev} in the conformal case, including
the hydrodynamic QNM frequency $\mathfrak{w}=-0.59806 i$. The presence of the dilaton field affects both the real and imaginary parts of
the QNMs, with the imaginary part generally showing greater sensitivity. For the hydrodynamic QNM, increasing the value of $\phi_{h}$
causes the imaginary part of the frequency to become less negative (closer to zero). This behavior is detailed in Table~\ref{Tab:HydroQNMVector},
where the numerical results for the hydrodynamic QNM are shown for selected values of $\mathfrak{q}$ and $\phi_{h}$.
For all displayed values of $\mathfrak{q}$, the imaginary part of the hydrodynamic frequency becomes progressively less negative as $\phi_{h}$ increases.

\section{The scalar sector of perturbations}\label{subsec:bulkviscosity}

\subsection{Master equation and hydrodynamic limit} 
Following the approach of Refs.~\cite{Gubser:2008sz,DeWolfe:2011ts}, an alternative procedure to calculate the bulk viscosity is to apply the Kubo formula to the conserved momentum flux (see Appendix \ref{app:KuboFormula} for details).
To simplify the analysis, we set the spatial component of the momentum $\mathfrak{q}$ to zero in the Fourier transforms of the perturbations that compose the scalar sector. This simplification results in three independent coupled differential equations for the perturbations $H_{tt}$, $H = H_{11}+H_{22}+H_{33}$, and $\chi$ (see Appendix \ref{Ap:scalar} for more details).

Using the gauge-invariant variable $Z_{\phi}$ [see Eq.~\eqref{Eq:APZphi}] and manipulating Eqs.~\eqref{Eq:ddhtt}--\eqref{Eq:dhii}, the following master differential equation is obtained,
\begin{align}\label{Eq:hseq}
    Z_{\phi}''& +  \left[\frac{f'}{f}-\frac{\zeta_{1}'}{\zeta_{1}}-\frac{2 \zeta_{1}''}{\zeta_{1}'}+\frac{2 \phi ''}{\phi '}\right] Z_{\phi}'\nonumber \\
    &+ \left[\frac{\mathfrak{w} ^2}{f^2}-\frac{f' \zeta_{1}'}{f \zeta_{1}}+\frac{f' \zeta_{1}''}{f \zeta_{1}'}-\frac{f' \phi ''}{f \phi '}\right] Z_{\phi}=0\, .
\end{align}

As described in Ref.~\cite{DeWolfe:2011ts}, the conserved flux factor can be obtained from the coefficient of $Z_{\phi}'$ in \eqref{Eq:hseq}.
As shown in Eq.~\eqref{Eq:FluxT}, the conserved flux factor is proportional to
\begin{equation}
\exp \left[  \int \left(\frac{f'}{f}-\frac{\zeta_{1}'}{\zeta_{1}}-\frac{2 \zeta_{1}''}{\zeta_{1}'}+\frac{2 \phi ''}{\phi '}\right) du \right]
= \frac{f}{\zeta_1}\frac{ \phi'^2}{ \zeta_{1}'^2},
\end{equation}
and the corresponding conserved flux for this sector can be written as
\begin{equation}\label{Eq:FluxHs}
    \mathcal{F}=\frac{2f}{3\zeta_{1}}\left[\frac{ \phi'}{\zeta_{1}'}\right]^2 \text{Im}\left(Z_{\phi}^{*}Z_{\phi}'\right),
\end{equation}
where $Z_{\phi}$ is a solution of Eq.~\eqref{Eq:hseq}. 

Even considering the particular case $\mathfrak{w}=0$, the solution for $Z_{\phi}$ is not analytical, and then it is necessary to use numerical methods. The boundary conditions that we must impose are $Z_{\phi}(0)=1$ at the boundary and the infalling-wave condition at the horizon. First, we consider the condition at the horizon. The general solution to \eqref{Eq:hseq} is given in the form \eqref{GeneralForm} with the exponent $\beta$ given by $\beta= i \omega/f'_{h}$. This choice from the two possible solutions $\beta = \pm i \omega/f'_{h}$ assures that we are selecting the infalling wave. Close to the horizon, it can be expressed as
\begin{equation}
     Z_{\phi}=\mathcal{C}_{\phi}(1-u)^{i \mathfrak{w}/f'_h} \label{Eq:Zphi_hh}
\end{equation}
where $\mathcal{C}_{\phi}$ is a normalization factor.
The second boundary condition, $Z_{\phi}(0)=1$, guarantees the correct extraction of the retarded Green's function and the bulk viscosity, since by the holographic dictionary it corresponds to turning on a unit source for the dual scalar operator. 

In the limit $\mathfrak{w}\to 0$, $Z_{\phi}$ approaches the constant $\mathcal{C}_{\phi}$. Substituting the approximation \eqref{Eq:Zphi_hh} into the expression for the conserved flux \eqref{Eq:FluxHs}, we obtain
\begin{equation}\label{Eq:fluxexpan}
   \mathcal{F}=\frac{2\mathfrak{w}}{3\zeta_{1h}} \left[\frac{ \phi'}{\zeta_{1}'}\right]_{u=1}^{2} |\mathcal{C}_{\phi}|^2\, .
\end{equation}
Finally, knowing the conserved flux, we can use the Kubo formula \eqref{Eq:ApBulkG} to obtain the bulk viscosity:
\begin{align}\label{Eq:BulkViscosity}
    \zeta\bigg|_{\text{GPR}}=
    \frac{8M_{p}^{3}N_{c}^{2}}{27\zeta_{1h}}\left[\frac{ \phi'}{\zeta_{1}'}\right]_{u=1}^{2} |\mathcal{C}_{\phi}|^2\,,
\end{align}
where GPR stands for Gubser, Pufu, and Rocha \cite{Gubser:2008sz}.

Consequently, the process of determining the bulk viscosity is reduced to evaluating the background solutions $\phi(u)$ and $\zeta_1(u)$, along with the constant $\mathcal{C}_{\phi}$. The constant $\mathcal{C}_{\phi}$ is extracted from the numerical solution of \eqref{Eq:hseq} with the appropriate boundary conditions. 

To ensure consistency, for the numerical calculation, we take $\mathfrak{w}=0$ from the beginning, and the initial conditions are derived from an expansion of $Z_{\phi}$ close to the horizon, besides imposing the infalling-wave condition for such a perturbation field. The expansion is implemented considering a small parameter $\delta$, defining an integration range from near the horizon up to a value $u_{min}=10^{-3}$ near the boundary, i.e., $1-\delta \geq u \geq u_{min}$, where $\delta=10^{-4}$. The expansion includes first- and second-order terms in $\delta$, thereby ensuring the expected behavior near the horizon. The coefficient $\mathcal{C}_{\phi}$ is obtained in an iterative way that ensures that the solution remains consistent.

\subsection{Bulk viscosity to entropy density ratio}

We can now use the GPR formula \eqref{Eq:BulkViscosity} and Eq.~\eqref{Eq:Entropy}
for $s$ to obtain the bulk viscosity to entropy density ratio:
\begin{equation}\label{Eq:GPRzeta_s}
    \frac{\zeta}{s}\bigg|_{\text{GPR}}=
    \frac{8}{27}\frac{\zeta_{1h}^{2}}{4\pi}\left[\frac{ \phi'}{\zeta_{1}'}\right]_{u=1}^{2} |\mathcal{C}_{\phi}|^2\, .
\end{equation}

Alternatively, the ratio $\zeta/s$ can be determined using a formula derived via the fluid/gravity correspondence:
\begin{equation}\label{Eq:formulaEO1}
    \frac{\zeta}{s}\bigg{|}_{\text{EO}}=
    \frac{8}{12\pi}s^{2}\left[\frac{d\phi_{h}}{ds}\right]^{2}
    = \frac{8}{12\pi}\left[\frac{d\phi_{h}/dz_{h}}{d\ln s/dz_h}\right]^{2},
\end{equation}
where EO stands for Eling and Oz \cite{Eling:2011ms}.
The entropy density is $s=4\pi M_{p}^{3}N_{c}^{2}/\zeta_{1h}^{3}$. So,
\begin{equation}
        \ln s = - 3 \ln\zeta_{1h} + \text{constant},
\end{equation}
where $\zeta_{1h} = \zeta_1(z=z_h)$. Using the result for $\zeta_1(z)$ given by Eq.~\eqref{Eq:ZetaAna} we obtain,
\begin{equation}
    \zeta_{1h} = z_h~{}_0F_{1}\left[\frac{5}{4},\frac{c^2z_h^4}{9}\right] . 
\end{equation}
From this expression and $\phi=cz^2=\phi_h u^2$, we see that
\begin{equation}
z_h\frac{d\ln \zeta_{1h}}{dz_h}=\frac{d\ln \zeta_1}{du}\bigg|_{u=1}
\quad\text{and}\quad z_h\frac{d\phi_{h}}{dz_{h}}=\frac{d\phi}{du}\bigg|_{u=1}.
\end{equation}
Substituting these results into \eqref{Eq:formulaEO1} yields 
\begin{equation}\label{Eq:EOeq}
    \frac{\zeta}{s}{\bigg{|}}_{\text{EO}}=
    \frac{8}{27}\frac{\zeta_{1h}^{2}}{4\pi}\left[\frac{\phi'}{\zeta'_{1}}\right]_{u=1}^{2}.
\end{equation}
Expression \eqref{Eq:EOeq} matches the result from the GPR formula \eqref{Eq:GPRzeta_s}, after making $|\mathcal{C}_{\phi}|^{2}=1$. 

\begin{figure}[ht!]
 \centering
 \includegraphics[width=7cm]{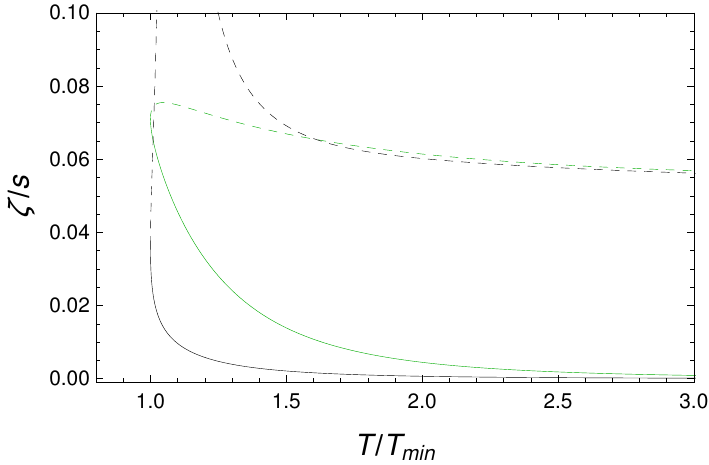}
 \caption{Bulk viscosity to entropy density ratio versus $T/T_{\text{min}}$. The values of $\zeta/s$ were obtained using the GPR formula (black line), with the numerically calculated $\mathcal{C}_{\phi}$, and the EO formula (green line), which corresponds to $|\mathcal{C}_{\phi}|=1$.}
 \label{Fig:BulkKubo}
\end{figure}

In Figure \ref{Fig:BulkKubo}, we show the bulk viscosity to entropy density ratio $\zeta/s$ considering Eqs.~\eqref{Eq:EOeq} (green line) and \eqref{Eq:GPRzeta_s} for $\mathcal{C}_{\phi}$ obtained numerically (black line).
In both curves, the solid lines represent the large black hole, while dashed lines represent the small black hole. We can observe that in the large black hole branch $\zeta/s$ continues to increase as $T$ decreases, up to the minimum temperature where the small black hole branch begins.   

The possibility of divergences of the ratio $\zeta/s$ along the branch of the small black hole has been mentioned in the literature for holographic models as in \cite{Gursoy:2009kk}. Also in Ref.~\cite{Gubser:2008yx}, it is established that $\zeta/s$ remains finite for $v_{s}^2 \geq 0$, which corresponds to the branch of large black holes, while for models that present regimes such that $v_{s}^2 \leq 0$, as in the present case, the ratio $\zeta/s$ may diverge.

\section{Comparison with data}
\label{Sec:Comparison}

In this section, we compare our results for the viscosities, primarily the bulk viscosity, with the result reported by the JETSCAPE collaboration  \cite{JETSCAPE:2020shq}, and we also compare the speed of sound with the Lattice QCD results \cite{Borsanyi:2012cr}.
For this comparison, we rewrite the coefficients in terms of the temperature (in MeV or GeV) and the dilaton parameter $c$ in $\text{GeV}^{2}$, focusing on the large black hole branch solutions. 

\begin{figure}[ht!]
 \centering
 \includegraphics[width=7cm]{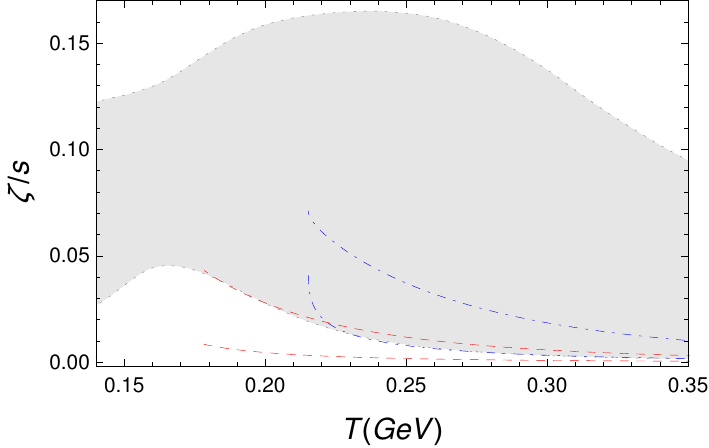}
 \caption{Bulk viscosity to entropy density ratio obtained using the Kubo formula for $c=0.22\,\text{GeV}^{2}$ (red dashed) and $c=0.4\,\text{GeV}^{2}$ (blue dot-dashed). The lower lines correspond to the numerically calculated $\mathcal{C}_{\phi}$, while the upper lines are obtained by putting $\mathcal{C}_{\phi}=1$; all cases in the large back hole branch. The dotted gray lines indicate the upper and lower bounds estimated by the JETSCAPE Collaboration}
 \label{Fig:BulkJET}
\end{figure}

Figure \ref{Fig:BulkJET} shows the comparison of our results for the bulk viscosity with the results obtained by the JETSCAPE Collaboration \cite{JETSCAPE:2020shq}, which are obtained from the analysis of experimental data on heavy-ion collisions.
The gray region bounded by dotted gray lines encloses the range of values of $\zeta/s$ reported by JETSCAPE. The dashed lines in color show results obtained in the present model for the large black hole branch. 
The lower set of lines represents our results for the ratio $\zeta/s$ obtained from the GPR formula \eqref{Eq:GPRzeta_s} with $\mathcal{C}_{\phi}$ calculated numerically. This solution matches the JETSCAPE range for $0.4\leq c \leq 1.0\,(\text{GeV}^{2})$ (blue dot-dashed lines).
The upper set of lines is obtained by fixing $\mathcal{C}_{\phi}=1$, which corresponds to the EO formula \eqref{Eq:EOeq}, and in this case matches the predictions of JETSCAPE for a wider range of values, specifically in the range $0.22\leq c \leq 1.0\,(\text{GeV}^{2})$ (dashed red lines).
Our results for $\zeta/s$ remain within the JETSCAPE uncertainty range across the relevant temperatures, suggesting that the holographic model employed here provides a reasonable description of the transport properties of the plasma.

Furthermore, the shear viscosity to entropy ratio $\eta/s$ remains constant at ($1/(4\pi)$), in agreement with the prediction from holographic QCD models. A comparison of this result with the JETSCAPE Collaboration data was made in Ref.\, \cite{Ballon-Bayona:2021tzw}.

\begin{figure}[ht!]
    \centering
    \includegraphics[width=7cm]{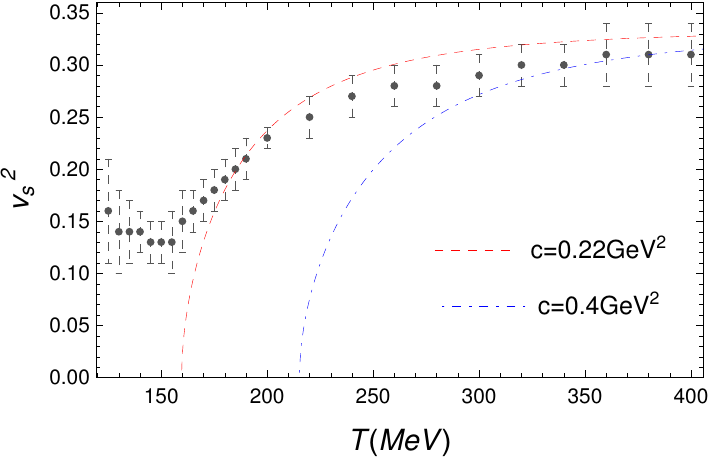}
    \caption{Speed of sound comparison between Lattice QCD result from Ref. \cite{Borsanyi:2012cr} (gray points) and our holographic model for $c=0.22 \,\text{GeV}^{2}$ (red dashed line) and  $c=0.4 \,\text{GeV}^{2}$ (blue dot-dashed line).}
    \label{Fig:SpeedLattice}
\end{figure}

In addition to the viscosity analysis, we examine the behavior of the speed of sound as a test of the thermodynamic consistency of the model. We compare our results for the squared speed of sound given by \eqref{Eq:SpeedSoundTherm} with results from Lattice QCD at zero chemical potential from Ref \cite{Borsanyi:2012cr}. Figure \ref{Fig:SpeedLattice} shows the comparison.

The gray points indicate the Lattice QCD results from Ref. \cite{Borsanyi:2012cr}, while the blue dotted-dashed and red dashed lines correspond to our results for the dilaton parameters $c=0.22\,\text{GeV}^{2}$ and $c=0.4\,\text{GeV}^{2}$, respectively. These values are consistent with the minimum values of $c$ for which our bulk viscosity results match the JETSCAPE data.
At low temperatures, our model does not reproduce the behavior shown in the Lattice QCD data. However, for $c=0.22\,\text{GeV}^{2}$, which has a minimum temperature of $178\, \text{MeV}$ close to $150\,\text{MeV}$, the match improves. This supports the consistency of our model in describing both transport and thermodynamic properties in the deconfined phase.

\section{Discussion and Conclusions}
\label{Sec:conclusion}

We have presented above the main results of a detailed study we conducted on an Einstein-dilaton model with a quadratic dilaton profile, 
focusing on the transport coefficients and investigating the influence of the dilaton field on such coefficients as compared
to the corresponding values obtained in the conformal limit, i.e., in the absence of the dilaton. To calculate the shear and bulk viscosities of this system within the framework of the AdS/QCD duality, we employed three complementary approaches, as summarized below. 

Using the hydrodynamic limit, following Ref.~\cite{Kovtun:2005ev}, we showed that no hydrodynamic mode exists in the tensor (traceless and transverse) sector. Meanwhile, in the vector sector, we successfully obtained the dispersion relation associated with the shear viscosity, obtaining the universal relation $\eta/s = 1/4\pi$. Furthermore, considering the entropy density $s$, see Eq.~\eqref{Eq:Entropy}, we derived the shear viscosity as a function of the temperature, as illustrated in Fig.~\ref{Fig:ShearViscosity}.

On the other hand, we presented a detailed numerical analysis of the spectrum of QNMs for the tensor and vector sectors. The relevant differential equations were derived from perturbations expressed in Eddington-Finkelstein coordinates. To compute the quasinormal frequencies, we employed a numerical package developed in Ref.~\cite{Jansen:2017oag}, which utilizes the pseudo-spectral method. Our numerical results are summarized in Table \ref{Tab:QNMsTT} for the tensor sector and in Tables~\ref{Tab:QNMsVectorial} and \ref{Tab:HydroQNMVector} for the vector sector.

The non-hydrodynamic modes in both sectors, shown in Fig.~\ref{Fig:QNMTTomegaVsq} for the tensor sector and Fig.~\ref{Fig:QNMvectorImomegaVsq} for the vector sector, exhibit similar behavior. The most significant modification occurs in the imaginary part of the frequency, indicating that the dilaton has a stronger influence for small $\mathfrak{q}$. As $\mathfrak{q}$ increases, the influence of the dilaton diminishes, and the behavior of the frequencies converges toward the conformal value, corresponding to $\phi_{h}=0$. In turn, the hydrodynamic mode of the vector sector also exhibited a dependence on the dilaton parameter, with its imaginary part becoming less negative as $\phi_{h}$ increases, as shown in Table \ref{Tab:HydroQNMVector} and Fig.~\ref{Fig:HydroMode}. Furthermore, we compared the analytical and numerical results for the dispersion relation and found good agreement, as illustrated in Fig.~\ref{Fig:ModoZeroShear}.

For the scalar sector, obtaining the bulk viscosity through the hydrodynamic limit posed significant challenges because to complications in solving the corresponding master equations. As an alternative, we applied the Kubo formula, following the procedure outlined in Ref.~\cite{DeWolfe:2011ts}. This approach allowed us to calculate the bulk viscosity and obtain the ratio $\zeta/s$ as a function of temperature. Our results are shown in Fig.~\ref{Fig:BulkKubo}, considering both the numerically computed integration constant $\mathcal{C}_{\phi}$ and the case where $\mathcal{C}_{\phi}=1$.
We verified that the bulk viscosity obtained via the Kubo formula matches the Eling-Oz relation when the normalization constant is set to $C_{\phi}=1$, confirming the consistency between the near-horizon dynamics and the hydrodynamic transport coefficients in the holographic model.
For $\mathcal{C}_{\phi}=1$, the ratio $\zeta/s$ reaches a maximum at $T_{min}$ and then approaches zero at high temperatures for large black holes. In the case of the numerically computed $\mathcal{C}_{\phi}$, the large black hole branch indicates that the bulk viscosity is higher near $T_{min}$ and also approaches zero at high temperatures.

Finally, the comparison between our results and data estimates from the JETSCAPE Collaboration, as well as the Lattice QCD simulations, demonstrates that our results capture key features of the strongly coupled quark-gluon plasma. The relationship between shear viscosity and entropy density $\eta/s=1/(4\pi)$, is within the JETSCAPE limits over most of the temperature range, as shown in Ref ~\cite{Ballon-Bayona:2021tzw}. The bulk viscosity $\eta/s$ is sensitive to non-conformal symmetry effects encoded in the dilaton profile, showing a significant temperature dependence that is in good agreement with the JETSCAPE uncertainty band. Furthermore, the squared speed of sound, calculated from the thermodynamics of black hole solutions, shows good agreement with Lattice QCD data at high temperatures, while deviations are expected at low temperatures. Together, these comparisons reinforce the applicability of these potential-reconstruction holographic models for describing the transport and thermodynamic properties of matter similar to QCD. 
We are interested in extending this model to reproduce the Lattice QCD results. As a natural next step, we are considering introducing a gauge field providing a framework for the transport coefficients description of plasmas at finite temperature and finite baryon density.

\acknowledgments

NAV is partially supported by Coordena\c{c}\~ao de Aperfei\c{c}oamento de Pessoal de N\'{i}vel Superior (CAPES), Brazil, Finance Code~001.
A.B-B acknowledges partial support from Conselho Nacional de Desenvolvimento Cient\'{i}fico e Tecnol\'{o}gico (CNPq, Brazil), Grant No.~314000/2021-6, and from CAPES, Brazil, Finance Code~001.
VTZ is supported by CNPq, Brazil, Grant No.~311726/2022-4.
This collaboration was also partially funded through a partnership between CAPES and FAPESB (Funda\c{c}\~ao Estadual de Amparo \`a Pesquisa do Estado da Bahia), within the scope of the \textit{Programa de Redu\c{c}\~ao de Assimetrias da P\'os-Gradua\c{c}\~ao} (PRAPG), through project 88887.909640/2023-00, entitled \textit{Planejamento estrat\'egico para o desenvolvimento do Programa de P\'os-Gradua\c{c}\~ao em F\'isica da UESC}.

\appendix

\section{Asymptotic analysis for large black holes}
\label{app:AsymptoticAnalysis}

Here we write the thermodynamic quantities for large black holes, i.e., small $z_h$ (or $u_h$). First, we expand the scale factor solution $\zeta_{1}(z)=z \, _0F_1\left[\frac{5}{4};\frac{c^2 z^4}{9}\right]$ around $z=0$, and evaluate in $z_{h}$ (or $u_{h}$). From this we can obtain the solution of $f(z\rightarrow 0)$ and the Hawking temperature, namely, 
\begin{equation}
    T(z_{h})\thickapprox \frac{1}{\pi  z_{h}}+\frac{2 c^2 z_{h}^3}{15 \pi }, \quad\quad z_{h}(T)\thickapprox \frac{1}{\pi T}+\frac{2 c^2}{15 \pi^5 T^5}\,.
    \label{Eq:zTUV}
\end{equation}
From Eqs.~\eqref{Eq:Entropy}, \eqref{Eq:ZetaAsyntotic}, and \eqref{Eq:zTUV} it is possible to obtain the behavior of entropy density as a function of temperature. In fact, expanding the entropy \eqref{Eq:Entropy} around $z=0$ and reversing the obtained series, we get 
\begin{equation}
    s(z_h)\thickapprox \frac{1}{2\pi z_h^3}-\frac{2 c^2 z_h}{15 \pi},
    \quad s(T)\thickapprox \frac{\pi^2 T^3}{2}-\frac{2 c^2}{15 \pi^2 T}\,.
    \label{Eq:suv}
\end{equation} 
As a function of temperature, the entropy increases until it reaches a constant value that corresponds to the conformal value at the boundary. 

The asymptotic forms of other relevant thermodynamic quantities, such as free energy $F(T)$, energy density $\epsilon(T)$, specific heat $C_{V}(T)$, and sound speed $v_{s}^2(T)$, are presented below.
\begin{align}
F(T)&\approx - \frac{\pi^2 T^4}{8}+\frac{2 c^2 }{15 \pi^2}\log (T);\label{FTUV}\\
\epsilon(T)&\approx \frac{3\pi ^2 T^4}{8}+ \frac{2c^2 \log (T)}{15 \pi ^2}-\frac{2 c^2}{15 \pi ^2};\label{ENEUV}\\
C_{V}(T)&\approx \frac{3 \pi^2 T^3}{2}+\frac{2 c^2}{15 \pi ^2 T};\\
v_{s}^2(T)&\approx \frac{1}{3}-\frac{16 c^2}{3 \left(4 c^2+45 \pi^4 T^4\right)}.
\end{align}
The results in this limit are compatible with the results obtained in the study of a conformal plasma, as expected, with an additional term due to the presence of the dilaton.

\section{Gauge-invariant variables}
\label{app:Invariant}

Under the infinitesimal coordinate transformations defined in \ref{Sec:gaugeInvQuantities}, the metric perturbations defined in Eq.~\eqref{Eq:Hsdefine} and the dilaton perturbation defined in Eq.~\eqref{Eq:dilatonpert} change as
\begin{align}
& H_{tt}\to H_{tt}+ \left(\frac{2 \partial_{z}\zeta_{1}}{\zeta_{1}}-\frac{\partial_{z}f}{f}\right)\varepsilon^{z} +2 i  \omega\varepsilon^{t}, \label{Eq:deltatt}\\
&H_{tj} \to  H_{t j} -i q\, \delta_{j 3}  f\varepsilon^{t}-i  \omega \varepsilon^{j},\label{Eq:deltatxi}\\
& H_{12}\to  H_{12},\label{Eq:deltax1x2}\\
& H_{j j}\to  H_{j j}+  i  q  \delta_{j 3}\,\varepsilon^{j}-\frac{2 \partial_{z}\zeta_{1}}{\zeta_{1}}\varepsilon^{z},\label{Eq:deltaxixi} \\ 
& H_{a 3}\to H_{a3} + i q\varepsilon^{a}
\, ,\label{Eq:deltaxix3}\\ 
&\,\chi \to \chi + \varepsilon^{z}\partial_{z}\phi\, \label{Eq:deltaphi},
\end{align}
where the indices $a,\, b$ run from $1$ to $2$, while $j$ runs from $1$ to $3$.

\subsection{Tensor sector} \label{Ap:tensor}

In the gauge adopted in the present work, the tensor sector is composed by just one perturbation field, i.e., the quantity $H_{12}$. 

In this sector, we find just one independent equation that may be written as
\begin{equation}
\begin{split}
    &\partial_{z}^{2}H_{12}-\partial_{z}\ln  \left[\frac{\zeta_{1}^4}{f \zeta_{2}}\right] \partial_{z}H_{12} +\frac{\zeta_{1}^2  \left({\omega}^2-q^2 f\right)}{\zeta_{2}^2~f^2 }H_{12}=0.
    \label{Eq:Hxy}
\end{split}
\end{equation}

Moreover, we observe that Eq.~\eqref{Eq:deltax1x2} already defines a gauge-invariant variable, and also note that it does not couple to any other perturbation field. We then rename $H_{12}$ as 
\begin{equation}\label{Eq:Zscalar}
    Z_{T}= H_{12}\, .
\end{equation}
Substituting this variable into \eqref{Eq:Hxy} and performing the normalization defined in Eq.~\eqref{Eq:normalization} we obtain the master equation \eqref{Eq:ZTmaster}.

\subsection{Vector sector } \label{Ap:vector}

This sector is composed by four perturbation fields, $H_{ta}$ and $H_{a3}$, where $a=1,\,2$, and there are six equations in total. Here, we write all of them.
\begin{align}
&  \partial_{z}^{2}H_{t a}-\partial_{z}\ln \left[\frac{\zeta_{1}^4}{\zeta_{2}}\right]\partial_{z}H_{t a}-\frac{q\zeta_{1}^{2}}{f\zeta_{2}^{2}}\left(q H_{t a} +\omega H_{a 3}\right) =0,\label{Eq:vec1}\\
&  \partial_{z}^{2}H_{a 3}-\partial_{z}\ln \left[\frac{\zeta_{1}^4}{f\zeta_{2}}\right] \partial_{z}H_{a 3} +\frac{\omega\zeta_{1}^{2}}{f^{2}\zeta_{2}^{2}}\left(q H_{t a} +\omega H_{a 3}\right)=0. \label{Eq:vec3}\\ 
&  \omega  \partial_{z}H_{t a}+q f \partial_{z}H_{a 3}=0. \label{Eq:vec2}
\end{align}
We can show that only four of these six equations are independent equations. 

In order to obtain the gauge-invariant variable of the vector sector, we combine \eqref{Eq:deltatxi} and \eqref{Eq:deltaxix3}
making \eqref{Eq:deltatxi}$/i \omega +$\eqref{Eq:deltaxix3}$/iq$ to eliminate $\epsilon^{a}$, i.e.,  
\begin{align}
& \delta H_{ta}=-i\omega \varepsilon^{a}\quad \longrightarrow \quad  \frac{i \delta H_{t a}}{\omega}=\varepsilon^{a}, \nonumber \\ 
& \delta H_{a 3}=i \,  q \,\varepsilon^{a} \,\quad \longrightarrow \quad - \frac{i\delta H_{a 3}}{q}=\varepsilon^{a},\nonumber \\ 
&\longrightarrow \delta \frac{i H_{t a}}{\omega}= \delta \frac{-i H_{a 3}}{q}\longrightarrow \delta\left(\frac{i H_{t a}}{\omega} + \frac{i H_{a 3}}{q}\right)= 0. \nonumber 
\end{align}
Hence,
\begin{equation}
Z_{V}= q H_{t a}~ +~\omega H_{a 3} \label{deltav}
\end{equation}
is a gauge-invariant variable.

A possible route to find the master equation \eqref{Eq:masterVector} is as follows.  
Write $H_{t a}$ in terms of $Z_V$ and $H_{a3}$ by using \eqref{deltav}, substitute the result into \eqref{Eq:vec2} to get an equation for $H_{a3}$ in terms of $Z_V$ alone, and then an equation for $H_{ta}$ in terms of $Z_V$ alone.  Finally, substituting these two relations into \eqref{Eq:vec1}, or into \eqref{Eq:vec3}, then making $\zeta_1=\zeta_2$, and eliminating $\zeta_1$ in terms of $f$ by using the background equations presented in Sec.~\ref{Sec:holographicmodel}, furnishes the final master equation for $Z_V$.
This procedure explicitly shows that Eqs.~\eqref{Eq:vec1} and \eqref{Eq:vec2} are not independent relations.

\subsection{Scalar sector}\label{Ap:scalar}

As in the case of the previous sectors of perturbations, our first goal is to use gauge-invariant quantities to write the master equations. In principle, the scalar mode contains six independent perturbation fields, namely $H_{tt}$, $H_{t3}$, $H=H_{11}+H_{22}+H_{33}$, and $\chi$, all coupled in a set of differential equations. Even following the ideas of Refs.~\cite{Kovtun:2005ev,Jansen:2019wag,Abbasi:2020ykq,Jansen:2017oag,Springer:2008js,Mas:2007ng}, such a coupling makes the computation of the QNMs and the subsequent extraction of the dispersion relation a quite complicated task. Starting with the fact that the construction of the gauge-invariant variable involves simplifying $\varepsilon^{j}$, $\varepsilon^{t }$ and $\varepsilon^{z}$, resulting in not one but two gauge-invariant variables and therefore two coupled master equations, whose solution is infeasible.

For practical reasons, we consider the approach shown in \cite{DeWolfe:2011ts}, we write the master equation for this sector considering $q=0$. In such a case, using the equations \eqref{Eq:deltaxixi} and \eqref{Eq:deltaphi} we may construct the gauge-invariant variable $Z_{\phi}$ as follows,
\begin{eqnarray}
   &&  \delta\chi =  \partial_{z}\phi \varepsilon^{z}  \qquad\longrightarrow \quad \varepsilon^{z}= \frac{\delta \chi}{\partial_{z}\phi}, \nonumber \\
   &&  \delta H=-\frac{6 \partial_{z}\zeta_{1}}{\zeta_{1}}\varepsilon^{z}  \quad  \longrightarrow\quad  \varepsilon^{z}= -\frac{\zeta_1\delta H}{6 \partial_{z}\zeta_{1}}
   , \nonumber \\ 
  && \longrightarrow \frac{\zeta_1\delta H}{6 \partial_{z}\zeta_{1}} + \frac{\delta \chi}{\partial_{z}\phi}=0 \longrightarrow \delta\left(\frac{\zeta_1 H}{6 \partial_{z}\zeta_{1}} + \frac{ \chi}{\partial_{z}\phi}\right)=0. \nonumber
\end{eqnarray}
Hence, the quantity 
\begin{equation}
  Z_{\phi} = \frac{H }{3} ~+~ \frac{2 \partial_{z}\zeta_{1}}{\zeta_{1} \partial_{z}\phi }\,\chi     \label{Eq:APZphi}
\end{equation}
is a gauge-invariant variable.

Note that $Z_{\phi}$ corresponds to the variable $\mathcal{H}$ in \cite{DeWolfe:2011ts}.

Next we show all the differential equations for the perturbations that correspond to this sector, which are obtained from the linear perturbations of \eqref{Eq:EinsteinEquation0} and \eqref{Eq:KleinGordon0} in the radial gauge ($h_{\mu z}=0$). 
\begin{align}
&   \partial_{z}^{2}H_{tt}+\partial_{z}\ln\left[\frac{\zeta_{1}}{f^{1/2}}\right]\partial_{z}H - \partial_{z}\ln\left[\frac{\zeta_{1}^{5}}{f^{3/2}\zeta_{2}}\right]\partial_{z}H_{tt} \nonumber\\ 
&-\frac{\omega ^2 \zeta_{1}^2 }{f^2 \zeta_{2}^2}H +\frac{16}{9}  \left( \partial_{z}\phi \partial_{z}\ln{\left[\frac{\zeta_{1}^{4}}{\zeta_{2}f}\right]} -\partial_{z}^{2}\phi\right)\chi=0, \label{Eq:ddhtt}\\ 
&   \partial_{z}^{2}H-\partial_{z}\ln{\left[\frac{\zeta_{1}^{7}}{\zeta_{2}f}\right]}\partial_{z}H +\frac{\omega ^2 \zeta_{1}^2 }{f^2 \zeta_{2}^2}H+\partial_{z}\ln{\left[\zeta_{1}^{3}\right]} \partial_{z}H_{tt}\nonumber\\ &-\frac{16}{3} \left( \partial_{z}\phi \partial_{z} \ln{\left[\frac{\zeta_{1}^{4}}{\zeta_{2}f}\right]}-\partial_{z}^{2}\phi\right)\chi =0,  \label{Eq:ddhii}\\
&  \partial_{z}^{2}H_{tt}-\partial_{z}^{2}H- \partial_{z}\ln\left[\frac{\zeta_{1}^{2}}{f^{3/2}\zeta_{2}}\right]\partial_{z}H_{tt} \nonumber \\ 
& +\partial_{z}\ln{\left[\frac{\zeta_{1}^{2}}{f^{1/2}\zeta_{2}}\right]}\partial_{z}H -\frac{16}{3}  \partial_{z}\phi\, \partial_{z}\chi  \nonumber\\
&  +\frac{16}{9} \left( \partial_{z}\phi \partial_{z} \ln{\left[\frac{\zeta_{1}^{4}}{\zeta_{2}f}\right]}-\partial_{z}^{2}\phi\right)\chi  =0,  \label{Eq:ddhmm}\\  
& \partial_{z}^{2}\chi-\partial_{z}\ln{\left[\frac{\zeta_{1}^{4}}{\zeta_{2}f}\right]}\partial_{z}\chi+\left(\frac{3 \partial_{\phi\phi}V(\phi )}{8 f \zeta_{2}^2}+\frac{\omega ^2 \zeta_{1}^2 }{f^2 \zeta_{2}^2}\right)\chi \nonumber \\& +\frac{1}{2}  \left(\partial_{z}H-\partial_{z}H_{tt}\right)\partial_{z}\phi  =0,  \label{Eq:ddchi}\\ 
&\partial_{z} H-\frac{\partial_{z} f}{2 f} H+\frac{8}{3} (\partial_{z}\phi) \chi=0.  \label{Eq:dhii}
\end{align}
It can be verified that two of these five equations are redundant. 

A procedure for obtaining the master equation for $Z_{\phi}$ is as follows. Write $\chi$ in terms of $Z_{\phi}$ and $H$ by using \eqref{Eq:APZphi} and substitute it into Eqs.~\eqref{Eq:ddhii}, \eqref{Eq:ddchi}, and \eqref{Eq:dhii}.  
Then, use the three resulting equations to eliminate $H_{tt}$ and $H$ in terms of $Z_{\phi}$ to get an uncoupled differential equation for $Z_{\phi}$.
Finally, incorporating the background equations, we simplify the resulting equation and cast it into the form of the master equation \eqref{Eq:hseq}.

\medskip

\section{Kubo Formula}
\label{app:KuboFormula}

The Kubo formula for the bulk viscosity is given by  
\begin{equation}
    \zeta=-\frac{4}{9}\lim_{\omega \rightarrow 0}\frac{1}{\omega}\text{Im}G_{R}(\omega).
\end{equation}
Following the conventions of Refs. \cite{Gubser:2008sz,DeWolfe:2011ts}, this formula can be written as
\begin{equation}\label{Eq:ApBulkG}
    \zeta=\frac{2}{9\kappa^2}\lim_{\omega \rightarrow 0}\frac{1}{\omega} \mathcal{F},
\end{equation}
where $\mathcal{F}$ is the conserved flux and $\kappa^2=1/\left(2M_p^3N_c^2\right)$. In the prescription of Refs.~\cite{Son:2002sd,Kovtun:2005ev, Herzog:2002pc}, the imaginary part of a retarded Green function, which is associated with the two-point correlation functions of an operator $\mathcal{O}$ in the dual field theory, is proportional to the conserved flux $\mathcal{F}$: 
\begin{equation}\label{Eq:ApbulkG_F}
\text{Im}G_{R}(\omega,q=0)=-M_{p}^{3}N_{c}^{2}\mathcal{F}.
\end{equation}

In \cite{Gubser:2008sz,Ballon-Bayona:2021tzw}, the conserved flux is obtained by the second-order expansion of the action. In \cite{DeWolfe:2011ts}, the flux is obtained directly by analyzing the master equation from Abel's identity. In systems governed by second-order differential equations, Abel's identity shows that the Wronskian is conserved along a coordinate and, in this case, the flux is independent of the holographic coordinate. In a general and illustrative way, we will consider a differential equation of the form
\begin{equation}
    y''(x)+ p(x)y'+q(x)y=0,
\end{equation}
the Wronskian is given by $W(y,y^{*})=y(y^{*})'-y^{*}y'$, where $y(x)$ and $y^{*}$ are two independent complex solutions. The conserved flux is associated with the imaginary part of the Wronskian
\begin{equation}\label{Eq:FluxT}
    \mathcal{F}\equiv \exp\left(\int p(x)dx\right)\text{Im}[y^{*},y'], 
\end{equation}
up to an overall factor.
To calculate the flux it is necessary to know the solution $y$ of the differential equation.

\bibliographystyle{apsrev4-2}

\bibliography{ref}

\end{document}